# Clarifying multiple-tip effects on Scanning Tunneling Microscopy imaging of 2D periodic objects and crystallographic averaging in the spatial frequency domain.

Jack C. Straton, Peter Moeck, Bill Moon Jr., and Taylor T. Bilyeu

Department of Physics, Portland State University, Portland, OR 97207-0751, U.S.A.

Crystallographic image processing (CIP) techniques may be utilized in scanning probe microscopy (SPM) to glean information that has been obscured by signals from multiple probe tips. This may be of particular importance for scanning tunneling microscopy (STM) and requires images from a sample that is periodic in two dimensions. The image-forming current for multiple tips in STM is derived in a more straightforward manner than prior approaches. The Fourier spectrum of the current for *p4mm* Bloch surface wave functions and a pair of delta function tips reveals the tip-separation dependence of various types of image obscurations. In particular our analyses predict that quantum interference should be visible on a macroscopic scale in the form of bands quite distinct from the basket-weave patterns a purely classical model would create at the same periodic double STM tip separations. A surface wave function that models the essential character of highly (0001) oriented pyrolytic graphite (technically known as HOPG) is introduced and used for a similar tip-separation analysis. Using a bonding $H_2$ tip wave function with significant spatial extent instead of this pair of infinitesimal Dirac delta function tips does not affect these outcomes in any observable way. This is explained by Pierre Curie's well known symmetry principle. Classical simulations of multiple tip effects in STM images may be understood as modeling multiple tip effects in images that were recorded with other types of SPMs). Our analysis clarifies *why* CIP and crystallographic averaging work well in removing the effects of a blunt SPM tip (that consist of multiple mini-tips) from the recorded 2D periodic images and also outlines the limitations of this image processing techniques for certain spatial separations of STM mini-tips.

## 1. INTRODUCTION

Scanning probe microscopy (SPM) images are often degraded due to the interference of two (or more) protrusions on the probe tip (i.e. effective mini-tips on a blunt tip), as well as containing sample tilt errors, image bow and drift, and stepping errors that occur while scanning the tip in two dimensions (2D) over the sample surface. Averaging methods have long been used to remove scanning errors. There are also well establish techniques for straightening out keystone-shaped images that result from sample tilt and image drift, [1] and for the removal of image bow by z-flattening using least-squares higher-order polynomials to model this distortion [2]. But removing multiple-tip defects of SPM images has only recently been accomplished through the use of crystallographic image processing (CIP) techniques [3,4,5], which one may consider as being a kind of a crystallographic averaging in reciprocal space of the intensity of symmetry related features in direct space.

Crystallographic image processing originated with the transmission electron crystallography community to aid the extraction of structure factor amplitudes and phase angles from high-resolution phase contrast images of crystalline materials [6, 7, 8] within the weak phase object approximation. It is also used for the correction of these images for the effects of the phase contrast transfer function, two-fold astigmatism, beam tilt away from the optical axis of the microscope, and sample tilt away from low indexed zone axes. CIP has also been has been applied to scanning transmission electron microscopy (STEM) [9].

Since one may define 2D image-based crystallography independent of the source of the 2D patterns as being concerned with categorizing, specifying, and quantifying 2D spatially periodic, perfectly long-range ordered patterns [5], CIP is also a good term for procedures as applied to SPM images of 2D periodic objects.

This process consists of the application of a Fourier transform to the 2D digitized image (called Fourier analysis), detection of the most likely plane symmetry in reciprocal space, enforcement of this symmetry by averaging of the symmetry related Fourier coefficients to remove all kinds of degradations, and finally inverse-Fourier image reconstruction (called Fourier synthesis into direct space). Consider, for example an STM image, Figure 1a, that has been recorded from a 2D monolayer array of fluorinated cobalt phthalocyanine ($F_{16}CoPc$) molecules on highly (0001) oriented pyrolytic graphite (which is technically known as HOPG). This was done at 20 K at the Technical University of Chemnitz under ultra high vacuum (UHV) conditions with a tungsten tip (that was held at room temperature). Figure 1b shows the inverse-Fourier image reconstruction after p4mm symmetry enforcement (using the guidelines in Appendix B). [10] Finally, Figure 1c shows one likely layout of nine $F_{16}CoPc$ molecules. The detail apparent in Figure 1b after CIP is highly suggestive of this layout, a correspondence not at all apparent in the unfiltered original image Figure 1a.

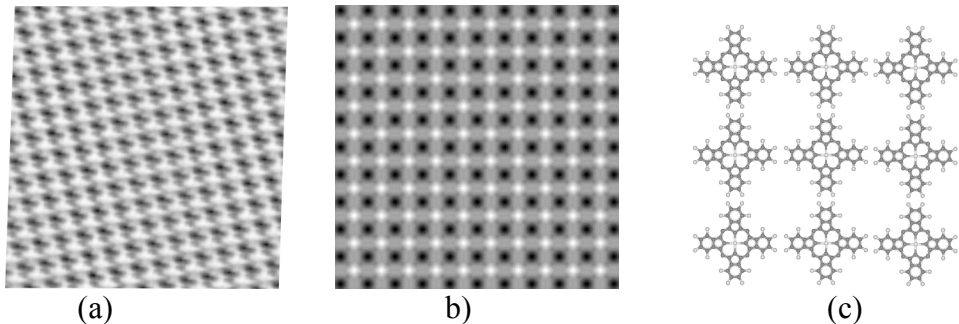

(a)                        b)                        (c)

**Figure 1.** A 2D monolayer array of $F_{16}CoPc$ molecules on HOPG **(a)** raw STM image, constant tunneling current mode, tip bias 1 V (with respect to the more negative

sample), 0.1 nA tunneling current; **(b)** *p4mm* plane symmetry enforced version of this data**;** and a model of nine molecules in one possible 2D periodic (*p4mm*) arrangement **(c)**. Modified after refs. [3] and [4].

The whole plane symmetry enforcing procedure can be thought of as aligning the periodic motifs of all independent STM images from the multiple mini-tips on top of each other, thus, enhancing the signal to noise ratio significantly when done correctly. The present work shows in detail *why* CIP works and builds upon prior work [5] that shows *how* it is done.

  We give a streamlined derivation of the tunneling current that the SPM images map. The changes wrought in the tunneling current by having two (or more) tips are then outlined. Next we show that the 2D Fourier transform of the derived current resulting from two tips is comprised of the same Fourier coefficients as a single tip, with the currents from the two tips differing only in a phase term in the momentum Fourier space. (CIP simply removes this phase term as part of the plane symmetry detection and enforcement procedures. There are, however, certain double tip separations for which some of these phases take certain Fourier coefficients to zero, thereby obstructing CIP.) The Fourier transform process is then explicated for a cosine Bloch wave function that gives an idealized approximation of the STM images that a real 2D-periodic sample with a square translational lattice and the comparatively high plane symmetry *p4mm* would produce, convolved with an idealized Dirac delta function wave function for each of the two tips. (We use the international (Hermann-Mauguin) notations for plane symmetry and 2D point symmetry groups [11] throughout the paper.) Analysis of the Fourier components shows at which tip separations one would expect to find both classical and quantum interference, a prediction borne out in the tunneling current images from this configuration of model surface and tip functions. It also examines the range of tip separations that are amenable to CIP.

  This process is repeated for a newly-derived, somewhat more complicated (0001) oriented graphite atomic surface model. Finally, as a check, Appendix A derives the current produced by a double tip with spatial extent, the bonding state of the $H_2$ molecule, with both the graphite and *p4mm* surface models and shows them to be proportional to the Dirac delta function cases. This expected outcome is explained therein by the application of Pierre Curie's well known symmetry principle.

  Our simulations are done without regards to any particulars of an actual scanning probe microscope. The classical simulations of STM images do not allow for interference between the two tips that constitute some kind of a blunt tip. One may, therefore, generalize the results of the classical simulations to other types of SPM images where there is no interference between the two tips.

## 2. The tunneling current

Since the SPM image is a two-dimensional map of the strength of the tunneling current density, we seek an expression for that quantity. Since the Hamiltonian of the full system

$$H = -\frac{\hbar^2}{2m}\nabla^2 + V \qquad\qquad (1a)$$

is generally too hard to solve, we follow Bardeen [12] and divide space into a barrier region (which may be of zero extent), a sample region, and a tip region. Then we define a sample potential V_S that is equal to V for points inside the sample and barrier regions and equal to 0 in the tip region (see Figure 2). The corresponding Hamiltonian H_S

$$H_s = -\frac{\hbar^2}{2m}\nabla^2 + V_s \quad (1b)$$

has surface eigenenergies and eigenfunctions labeled by quantum numbers σ. Likewise we have V_T = V for points inside the tip and barrier regions, and equal to zero in the sample region. The corresponding Hamiltonian H_T

$$H_T = -\frac{\hbar^2}{2m}\nabla^2 + V_T \quad (1c)$$

has tip eigenenergies and eigenfunctions are labeled by quantum numbers τ.

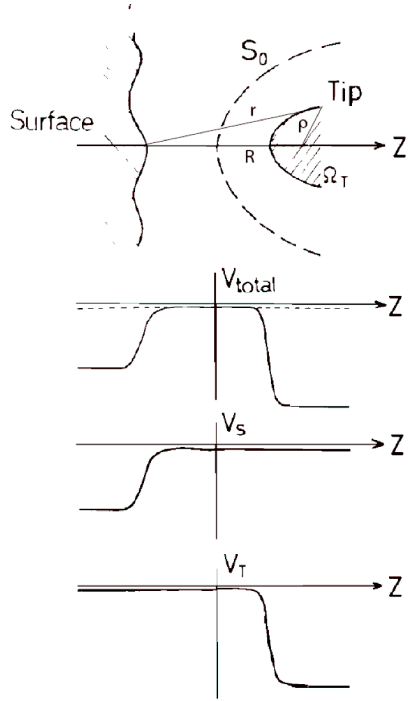

**Figure 2.** Adapted from Tsukada and Shima [13].

The matrix element for an electron initially in the surface state $\psi_\sigma$ to scatter into the tip state $\psi_\tau$ -- also the matrix element for an electron initially in the tip state $\psi_\tau$ to scatter into the surface state $\psi_\sigma$ since we will be taking the absolute square -- is approximately [14]

$$M_{\sigma\tau} = \langle \sigma | H - H_S | \tau \rangle = \int d^3 r \psi_\sigma^* [H - H_S] \psi_\tau = \int_{\Omega_T} d^3 r \psi_\sigma^* [H - H_S] \psi_\tau = \int_{\Omega_T} d^3 r \psi_\sigma^* [V - V_S] \psi_\tau$$

(2a).

The third equality acknowledges that on the sample side of the surface the quantity H − H_S = V - V_S is zero, so we can confine the integral to the region of the tip. Bardeen goes on to make this matrix element more symmetrical with respect to wave functions, and transforms it into an integral over the surface $S_0$ that encloses the volume $\Omega_T$ by using Green's first identity [15], $E_\sigma = E_\tau$, and (1b):

$$M_{\sigma\tau} = \frac{\hbar^2}{2m} \int\limits_{S_0} d\mathbf{S} \cdot \left( \psi_\sigma^* \nabla \psi_\tau - \psi_\tau \nabla \psi_\sigma^* \right) \tag{2b}.$$

As the volume integral form is a more visually intuitive form for many purposes, Tsukada and Shima [13] reverse Bardeen's derivation, starting with the symmetrical matrix element $M_{\sigma\tau}$ (1b). It is more convenient, however, to simply start with (2a) and subtract from it a quantity, $H - H_T = V - V_T$, that is the zero operator on the tip side of the surface [14];

$$0 = \int\limits_{\Omega_T} d^3 r \psi_\sigma^* \left[ H - H_T \right] \psi_\tau = \int\limits_{\Omega_T} d^3 r \psi_\sigma^* \left[ V - V_T \right] \psi_\tau \tag{3}.$$

Then

$$M_{\sigma\tau} = -\int\limits_{\Omega_T} d^3 r \psi_\sigma^* \left[ V_S - V_T \right] \psi_\tau \tag{2c},$$

which is proportional to Equation (A-3) in Tsukada and Shima [13]. The following term is the core of their tunneling current [13]

$$I = \frac{2\pi e}{\hbar} \sum_{\sigma\tau} \left[ f\left( E_\sigma \right) - f\left( E_\tau - eV \right) \right] \left| M_{\sigma\tau} \right|^2 \delta\left( E_\sigma - E_\tau \right) \tag{4a},$$

an expression of the Dirac-Wentzel-Fermi Golden Rule [16].

Finally we introduce another identity in (4a)

$$\delta\left( E_\sigma - E_\tau \right) = \int_{-\infty}^\infty dE \delta\left( E - E_\sigma \right) \delta\left( E - E_\tau \right) \tag{5},$$

and take the zero-temperature limit [17] to give

$$I = \frac{2\pi e}{\hbar} \int_{E_F}^{E_F + eV} dE \left[ f(E) - f(E - eV) \right] A\left( \mathbf{R}, E, E + eV \right) \tag{4b},$$

where

$$\begin{aligned}
A\left( \mathbf{R}, E, E' \right) &= \pi^2 \int\limits_{\Omega_T} d^3 r \int\limits_{\Omega_T} d^3 r' \left[ V_S(\mathbf{r}) - V_T(\mathbf{r} \cdot \mathbf{R}) \right] \left[ V_S(\mathbf{r}') - V_T(\mathbf{r}' \cdot \mathbf{R}) \right] \\
&\quad \times g^S\left( \mathbf{r}', \mathbf{r}, E \right) g^T\left( \mathbf{r} \cdot \mathbf{R}, \mathbf{r}' \cdot \mathbf{R}, E' \right) \\
&= \pi^2 \int\limits_{\Omega_T} d^3 \rho \int\limits_{\Omega_T} d^3 \rho' V_T(\rho) V_T(\rho') \, g^S\left( \rho' + \mathbf{R}, \rho + \mathbf{R}, E \right) g^T\left( \rho, \rho', E' \right)
\end{aligned} \tag{6a},$$

in which $g^S$ represents a sum over conjugate pairs of surface eigenfunctions $\psi_\sigma$,

$$g^S\left( \rho' + \mathbf{R}, \rho + \mathbf{R}, E \right) \equiv \sum_\sigma \psi_\sigma\left( \rho' + \mathbf{R} \right) \psi_\sigma^*\left( \rho + \mathbf{R} \right) \delta\left( E - E_\sigma \right) \tag{7a}.$$

Likewise for the tip,

$$g^T\left( \rho, \rho', E \right) \equiv \sum_\tau \psi_\tau\left( \rho \right) \psi_\tau^*\left( \rho' \right) \delta\left( E - E_\tau \right) \tag{7b}.$$

As a number of authors note, each $g$ can be represented as ($\pi$ times) the imaginary part of a Green's function [18]. In the last expression of (6a) we have removed $V_S$ since it is negligible in the region on the tip side of the surface $S_0$, which can be taken as close to the tip surface as we like [13].

Tsukada and Shima [13] show that (4b) and (6a) reduce to essentially the result of Tersoff and Hamann [19] with their choice of tip function but with the latter's R — defined as the weighted center of curvature of the tip — replaced by the average center of mass of the product of tip potential, tip eigenfunction, and a factor characterizing the decay of the surface function into the deeper region of the tip.

## 3. The two-tip tunneling current

Extending this result to encompass two-tips is now straightforward. We follow the derivation of Tsukada, Kobayashi, and Ohnishi [20], though in more detail. Suppose we split the tip integration into two regions to reflect a double tip as in Figure 3.

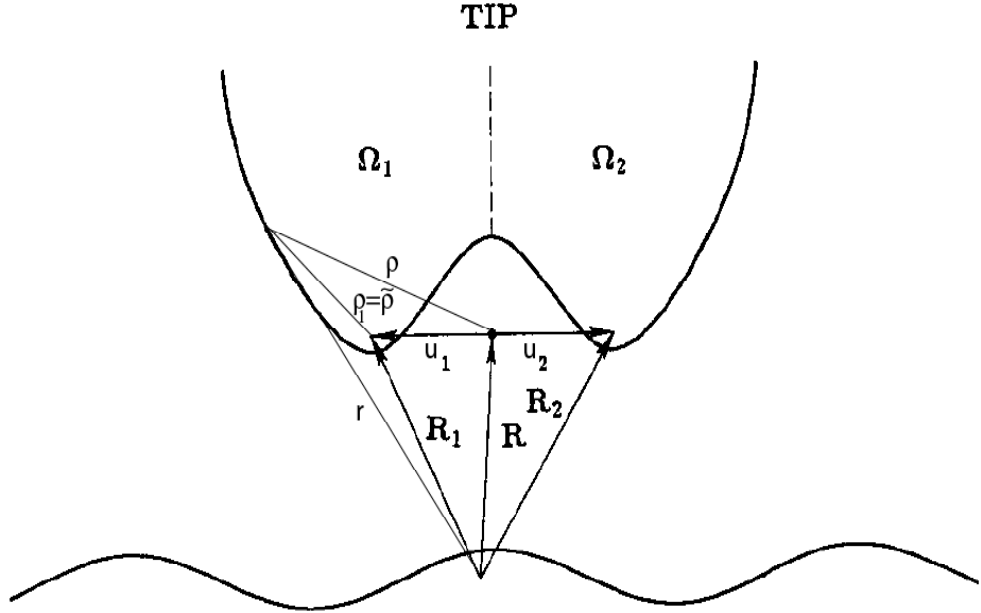

**Figure 3.** Adapted from Tsukada, Kobayashi, and Ohnishi [20].

Then

$$M_{\sigma\tau} = \int_{\Omega_T} d^3\rho \left( \psi_\sigma^* \left[ V_S - V_T \right] \psi_\tau \right) = \int_{\Omega_1} d^3\rho \left( \psi_\sigma^* \left[ V_S - V_T \right] \psi_\tau \right) + \int_{\Omega_2} d^3\rho \left( \psi_\sigma^* \left[ V_S - V_T \right] \psi_\tau \right) \quad (2d).$$

If we change variables in the first integral to $\boldsymbol{\rho} = \boldsymbol{\rho}_1 + \mathbf{u}_1$ ($\mathbf{u}_1$ is the relative coordinate to the center of mass $\mathbf{R}_1$ of the half-tip), and likewise for integration region $\Omega_2$, and then rename $\boldsymbol{\rho}_m$ as $\tilde{\boldsymbol{\rho}}$ in both cases to reduce notational complexity in what follows, then (6a) becomes

$$A(\mathbf{R}, E, E') = N(1, \mathbf{R}, E, E') + N(2, \mathbf{R}, E, E') + I(1, 2, \mathbf{R}, E, E') + I(2, 1, \mathbf{R}, E, E') \quad (6b),$$

where the first two terms represent sources in which the current follows two independent classical paths,

$$N\left(m,\mathbf{R},E,E'\right)=\int_{\Omega_i}d^3\tilde{\rho}\int_{\Omega_i}d^3\tilde{\rho}'V_T\left(\tilde{\boldsymbol{\rho}}+\mathbf{u}_m\right)V_T\left(\tilde{\boldsymbol{\rho}}'+\mathbf{u}_m\right)$$

$$\times g^S\left(\tilde{\boldsymbol{\rho}}'+\mathbf{R}_m,\tilde{\boldsymbol{\rho}}+\mathbf{R}_m,E\right)g^T\left(\tilde{\boldsymbol{\rho}}+\mathbf{u}_m,\tilde{\boldsymbol{\rho}}'+\mathbf{u}_m,E'\right)$$

(8a).

The last two terms represent the interference between two quantum paths for each electron,

$$I\left(m\neq n,\mathbf{R},E,E'\right)=\int_{\Omega_1}d^3\tilde{\rho}\int_{\Omega_2}d^3\tilde{\rho}'V_T\left(\tilde{\boldsymbol{\rho}}+\mathbf{u}_m\right)V_T\left(\tilde{\boldsymbol{\rho}}'+\mathbf{u}_n\right)$$

$$\times g^S\left(\tilde{\boldsymbol{\rho}}'+\mathbf{R}_m,\tilde{\boldsymbol{\rho}}+\mathbf{R}_n,E\right)g^T\left(\tilde{\boldsymbol{\rho}}+\mathbf{u}_m,\tilde{\boldsymbol{\rho}}'+\mathbf{u}_n,E'\right)$$

(9a).

Equations (6b) through (9a) are equivalent to the final result of Tsukada, Kobayashi, and Ohnishi [20].

Mizes, Park, and Harrison [21] also examined the two-tip problem. They hypothesized that the relative interference phases of many tips would tend to sum to zero, which seems reasonable. But their application of this conclusion to just two tips is logically inconsistent: here there is just one relative phase that cannot sum itself to zero.

Indeed, Tsukada, Kobayashi, and Ohnishi [20] calculate the interference of a double tip over graphite and show that "banding" in their numerically-generated STM image can result when the interference term (9a) is large.

## 4. The Fourier Spectrum

To examine the utility of CIP for the double tip case, let

$$\psi_\sigma\left(\mathbf{r}\right)\equiv\Psi_\sigma\left(x,y\right)\Phi_\sigma\left(z\right)$$

(10),

whose first factor is a continuous periodic function in the plane of the surface. It thus has an X-Y variation that can be represented by a 2D Fourier transform whose argument is a discrete set of complex numbers:

$$\Psi_\sigma\left(x,y\right)^{\text{periodic}}=\frac{1}{2\pi}\int_{-\infty}^{\infty}\int_{-\infty}^{\infty}F_\sigma\left(P,Q\right)\exp\left[-i\left(xP+yQ\right)\right]dPdQ$$

$$=\frac{1}{2\pi}\sum_{-P}^{P}\sum_{-Q}^{Q}F_\sigma\left(P,Q\right)^{\text{discrete}}\exp\left[-i\left(xP+yQ\right)\right]$$

(11)

The product of such wave functions in $g^S$ of Equation (7a) is the only part of the current source that depends on the external tip position variable R through $\mathbf{R_m}=\mathbf{R}+\mathbf{u}_m$. Then the Fourier transform of $g^S$ with respect to $\mathbf{R}$ is given by the convolution,

$$G^S\left(h,k,(\boldsymbol{\rho}+\mathbf{R}+\mathbf{u_m})\cdot\hat{\mathbf{e}}_3,(\boldsymbol{\rho}'+\mathbf{R}+\mathbf{u_m})\cdot\hat{\mathbf{e}}_3,E\right)$$

$$=\frac{2\pi}{\pi}\sum_\sigma \Phi_\sigma\left((\boldsymbol{\rho}+\mathbf{R}+\mathbf{u_m})\cdot\hat{\mathbf{e}}_3\right)\Phi_\sigma^*\left((\boldsymbol{\rho}'+\mathbf{R}+\mathbf{u_m})\cdot\hat{\mathbf{e}}_3\right)\delta(E-E_\sigma)$$

$$\times\sum_{-P}^{P}\sum_{-Q}^{Q}F_\sigma(P,Q)\exp\left\{-i\left[(\boldsymbol{\rho}+\mathbf{u_m})\cdot\hat{\mathbf{e}}_1 P+(\boldsymbol{\rho}+\mathbf{u_m})\cdot\hat{\mathbf{e}}_2 Q\right]\right\}$$

$$\times F_\sigma^*\left((p-P),(q-Q)\right)\exp\left\{i\left[(\boldsymbol{\rho}'+\mathbf{u_m})\cdot\hat{\mathbf{e}}_1(p-P)+(\boldsymbol{\rho}'+\mathbf{u_m})\cdot\hat{\mathbf{e}}_2(q-Q)\right]\right\}$$

$$=2\sum_\sigma \Phi_\sigma\left((\boldsymbol{\rho}+\mathbf{R}+\mathbf{u_m})\cdot\hat{\mathbf{e}}_3\right)\Phi_\sigma^*\left((\boldsymbol{\rho}'+\mathbf{R}+\mathbf{u_m})\cdot\hat{\mathbf{e}}_3\right)\delta(E-E_\sigma)$$

$$\times\sum_{-P}^{P}\sum_{-Q}^{Q}F_\sigma(P,Q)\exp\left\{-i\left[(\boldsymbol{\rho}-\boldsymbol{\rho}'+\mathbf{2u_m})\cdot\hat{\mathbf{e}}_1 P+(\boldsymbol{\rho}-\boldsymbol{\rho}'+\mathbf{2u_m})\cdot\hat{\mathbf{e}}_2 Q\right]\right\}$$

$$\times F_\sigma^*\left((p-P),(q-Q)\right)\exp\left\{i\left[(\boldsymbol{\rho}'+\mathbf{u_m})\cdot\hat{\mathbf{e}}_1 p+(\boldsymbol{\rho}'+\mathbf{u_m})\cdot\hat{\mathbf{e}}_2 q\right]\right\}$$

$$(7c).$$

Equation (7c) explicitly shows that the currents from the two tips (m = 1, 2) differ only in a phase term in the momentum Fourier space

$$\exp\left\{-i\left[(\mathbf{u_1}-\mathbf{u_2})\cdot\hat{\mathbf{e}}_1(2P-p)+(\mathbf{u_1}-\mathbf{u_2})\cdot\hat{\mathbf{e}}_2(2Q-q)\right]\right\} \qquad (12)$$

We will see that this phase term does not alter the configuration of frequencies of the Fourier transform we use in CIP, but can affect the magnitudes of the Fourier coefficients (that are related to the square roots of the image intensity) of the discrete points (P,Q).

This point may be most clearly demonstrated if we utilize the simplest possible wave functions that will reveal interference. We do so to illustrate the possibilities for reciprocal space symmetry enforcement, not to make predictions for any particular sample or tip. Thus, these wave functions will not be chosen to be eigenfunctions of the Hamiltonians $H_S$ or $H_T$.

The simplest possible double-tip wave function is a pair of infinitesimal tips, each of whose amplitude is zero everywhere outside of the tip centers of mass points $\mathbf{R_m}=\mathbf{R}+\mathbf{u_m}$, which is the point $\tilde{\boldsymbol{\rho}}=0$ (remembering that we renamed each $\boldsymbol{\rho}_m$ as $\tilde{\boldsymbol{\rho}}$)

$$g^T\left(\tilde{\boldsymbol{\rho}},\tilde{\boldsymbol{\rho}}',E'\right)\equiv\sum_\tau \psi_\tau(\tilde{\boldsymbol{\rho}})\psi_\tau^*(\tilde{\boldsymbol{\rho}}')\delta(E-E_\tau)\rightarrow\delta(\tilde{\boldsymbol{\rho}})\delta(\tilde{\boldsymbol{\rho}}')\delta(E'-E_T) \qquad (7d).$$

Then

$$N(m,\mathbf{R},E,E')=\int_{\Omega_i}d^3\tilde{\boldsymbol{\rho}}\int_{\Omega_i}d^3\tilde{\boldsymbol{\rho}}' V_T\left(\tilde{\boldsymbol{\rho}}+\mathbf{u_m}\right)V_T\left(\tilde{\boldsymbol{\rho}}'+\mathbf{u_m}\right)$$

$$\times g^S\left(\tilde{\boldsymbol{\rho}}'+\mathbf{R}+\mathbf{u_m},\tilde{\boldsymbol{\rho}}+\mathbf{R}+\mathbf{u_m},E\right)g^T\left(\tilde{\boldsymbol{\rho}}+\mathbf{u_m},\tilde{\boldsymbol{\rho}}'+\mathbf{u_m},E'\right) \qquad (8b).$$

$$\rightarrow V_T(\mathbf{u_m})V_T(\mathbf{u_m})g^S\left(\mathbf{R}+\mathbf{u_m},\mathbf{R}+\mathbf{u_m},E\right)\delta(E'-E_T)$$

This mathematically convenient, but physically unrealizable approximation of the tip wave functions shows that the two classical current sources are simply translated left or right relative to each other, so a Fourier transform of these two terms of the double tip differ only in a phase in reciprocal space.

The two terms that represent the interference between two quantum paths for each electron are somewhat more complicated,

$$I\left(m \neq n, \mathbf{R}, E, E'\right) = \int\limits_{\Omega_1} d^3\tilde{\rho} \int\limits_{\Omega_2} d^3\tilde{\rho}' V_T\left(\tilde{\rho} + \mathbf{u}_m\right) V_T\left(\tilde{\rho}' + \mathbf{u}_n\right)$$

$$\times\, g^S\left(\tilde{\rho}' + \mathbf{R} + \mathbf{u_m}, \tilde{\rho} + \mathbf{R} + \mathbf{u_n}, E\right) g^T\left(\tilde{\rho} + \mathbf{u}_m, \tilde{\rho}' + \mathbf{u}_n, E'\right) \quad (9b),$$

$$\rightarrow V_T\left(\mathbf{u}_m\right) V_T\left(\mathbf{u}_n\right) g^S\left(\mathbf{R} + \mathbf{u_m}, \mathbf{R} + \mathbf{u_n}, E\right) \delta\left(E' - E_T\right)$$

partly shifting leftward and partly rightward, but again will differ only in a phase in reciprocal space.

The Dirac delta function we have used to approximation the tip wave functions can be represented as the limit of the following Gaussian function as $\varepsilon \rightarrow 0$:

$$\frac{1}{2\sqrt{\pi\varepsilon}} e^{-\tilde{\rho}^2/(4\varepsilon)} \quad (13).$$

If we use a somewhat more realistic approximation of the tip wave functions by, say, utilizing (13) with a small but non-zero value for $\varepsilon$, the integrals in (8b) and (9b) would sample a small region of $g^S$ nearby $\mathbf{R} + \mathbf{u}_m$, giving us a somewhat smeared version of (8b) and (9b). While that would better model an actual experiment, it would make our examination of the effects of quantum interference less clear and would, hence, be less desirable for the purposes of illustrating crystallographic image processing (CIP). We do, however, show in appendix A that using a bonding $H_2$ tip wave function [22] with significant spatial extent, instead of this pair of infinitesimal Dirac delta function tips, does not affect the outcome in any observable way.

For our surface wave functions we again choose the simplest possible wave functions that will reveal the difference between the classical superposition of two imaging tips and the addition of quantum interference. Chen [23] uses a cosine Bloch wave function (his Equations (5.33) and (5.34)) that gives us an idealized approximation of the images that a real *p4mm* sample would produce,

$$\psi_1(x, y, z) = e^{-1/2 \gamma z}\left(\cos(x) + \cos(y)\right) \quad (14).$$

We first create current maps from only the superpositions of the two classical current sources. If the tips are separated in the x direction by 2u, Equation (8b) (with this real function (14)) will give a superposed current proportional to

$$\left[\psi_1(x-u, y, 1)\right]^2 + \left[\psi_1(x+u, y, 1)\right]^2 =$$

$$(2 + \text{Cos}[2\,u]\,\text{Cos}[2\,x] + 4\,\text{Cos}[u]\,\text{Cos}[x]\,\text{Cos}[y] + \text{Cos}[2\,y])\,/\,e$$

$$(15),$$

We have set z = $\gamma$ = 1 for convenience.

The Fourier transform of (15) is, term by term [24],

$$\mathcal{F}\left[\left[\psi_1(x-u, y, 1)\right]^2 + \left[\psi_1(x+u, y, 1)\right]^2\right] = \frac{\pi}{e}\left(4\,\delta[P]\,\delta[Q]\right.$$

$$+ \text{Cos}[2\,u]\left(\delta[-2+P] + \delta[2+P]\right)\delta[Q]$$

$$+ 2\,\text{Cos}[u]\left(\left(\delta[-1+P] + \delta[1+P]\right)\delta[-1+Q] + \left(\delta[-1+P] + \delta[1+P]\right)\delta[1+Q]\right)$$

$$+ \delta[P]\left(\delta[-2+Q] + \delta[2+Q]\right)$$

$$(16).$$

This confirms that classically the conjugate lattice spacing, and hence the underlying symmetry in real space, is independent of the number of tips. A double tip separated by 2u introduces phase terms, Cos[a u], whose only role is to suppress various frequencies at specific tip separations. When u = π/4, for instance, the points at (P,Q) = (±2,0) are suppressed, as are the points at (P,Q) = (±1,±1) when u = π/2.

   Rather than using reciprocal space Cartesian coordinates (P,Q) of our analytic Fourier transform above, crystallographers generally use lattice vector components {H,K}, that are in general non-orthogonal.  Thus, in the present case, the popular CIP program CRISP [25] automatically chooses a lattice vector {H,K} = {1,0} to go from the origin to the bright point at 45° to the (horizontal) positive Q axis in Figure 4 (a), which has orthogonal coordinates (P,Q) = (1,1). This is one of the points suppressed when u = 1.54, just short of π/2, in Figure 4(d).  CRISP chooses {H,K} = {0,1} to go from the origin to the bright point in Figure 4(a) at 135° to the positive Q axis, which has orthogonal coordinates (P,Q) = (-1,1).  In the crystallographic convention, the lattice vector sum {H,K} = {1,-1} corresponds to the point (P,Q) = (2,0) that is suppressed when u = π/4   in Figure 4(b), where we have taken u = 0.77 just short of π/4.

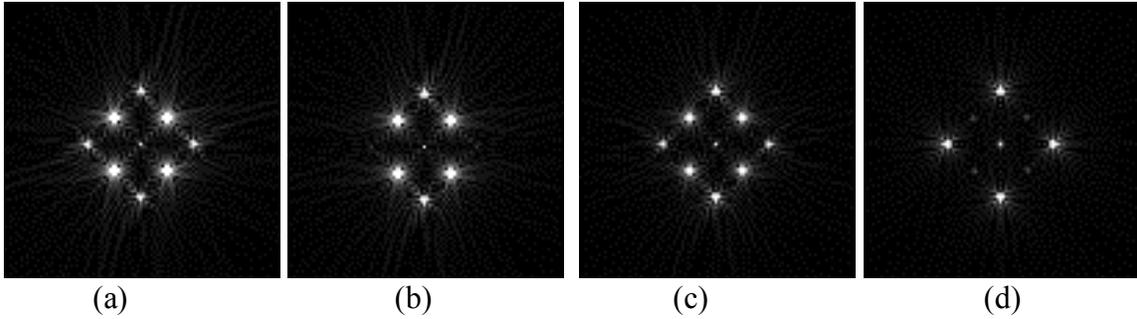

(a).                     (b).                     (c).                     (d).

**Figure 4.** Fast Fourier transforms of the square of the classical sum of idealized *p4mm* wave functions (15a) with (a) zero tip separation, (b) u = 0.77 = π/4 −ε, (c) u = 1.1, and (d) u = 1.54 = π/2 −ε.

This results in a significant change in the image registered by this model double STM tip, as seen in Figure 5(b). Even with such significant suppression of spatial frequency information, CIP still is able to recover an excellent symmetrized "image" of the classical sum of idealized *p4mm* wave functions, as seen in Figure 6(b), when compared with the single-tip image 5(a). Essentially the same applies to u = 1.1, Figure 5(c), as illustrated by the CIP symmetrized image of Figure 6(c).
(Appendix B gives first a brief introduction on how to determine (in a semi-quantitative manner) the plane symmetry to which an image most likely belongs and then analyses the plane symmetries in Figures 5(b) to 5(d) in order to lend support to the assertions we have just made.)

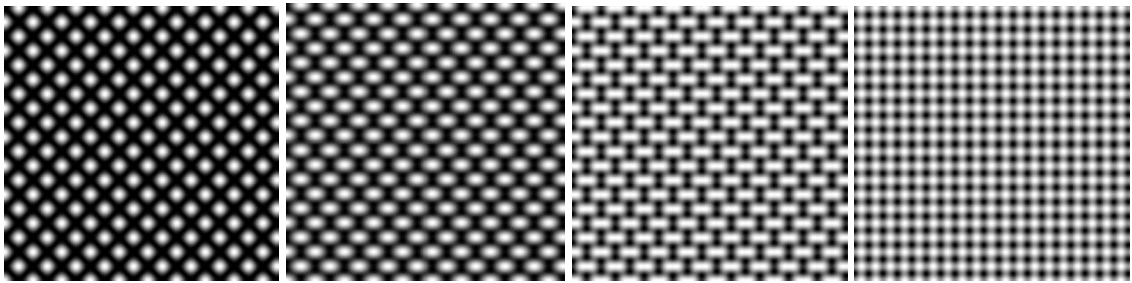

(a)                    (b)                    (c)                    (d)

Figure 5. Superpositions of the two classical current sources with delta function STM tip half-separation (a) u = 0.0, (b) u = 0.77 = π/4 − ε, (c) u = 1.1, and (d) u = 1.54 = π/2 − ε units in the horizontal direction, relative to the *p4mm* Bloch surface functions having a period of 2π.

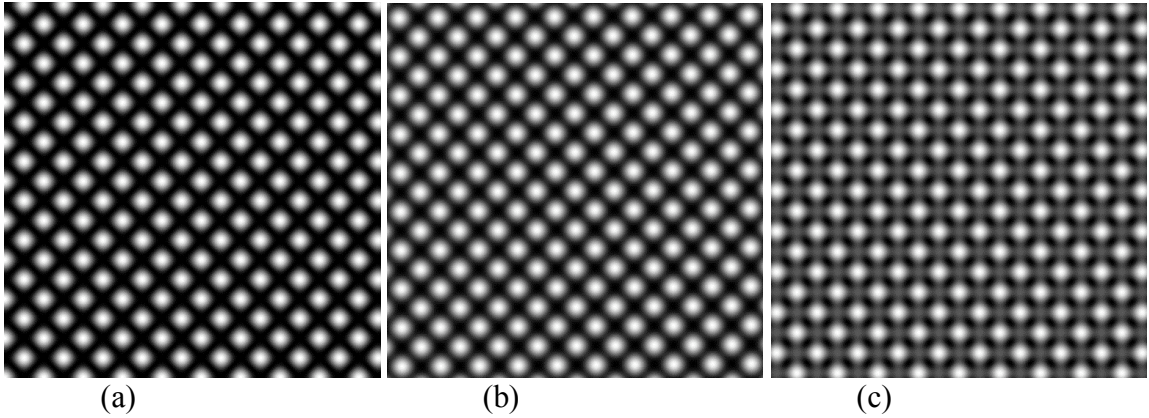

(a)                              (b)                              (c)

**Figure 6.** Plane symmetry enforcement of the underlying p4mm symmetry in reciprocal space for the superposition of the two classical current sources with delta function STM tip half-separation Figures 5. (a) u = 0.0, (b) u = 0.77 = π/4 − ε, and (c) u = 1.1 units in the horizontal direction, relative to the *p4mm* Bloch surface functions having a period of 2π.

On the other hand, with u = 1.54 = π/2 − ε the *p4mm* symmetry of the "sample" is entirely obscured, in Figure 5(d), by the suppression of the (P,Q) = (±1,±1) frequencies due to the classical superposition of these two images. This becomes clear when we examine the standard "plane symmetry quantifiers" for all non-hexagonal plane groups, as defined in appendix B.

So even if quantum interference were missing from the two-tip problem, as postulated by Mizes, Park, and Harrison [21], there remain severe classical imaging problems that can for result from certain double STM tip separations that are too large for CIP to deal with.

## 5. Quantum Interference

It is fascinating that an image recorded from two tips can be obscured by the suppression of the various frequencies due to the superposition of two classical images at specific tip separations. One would expect that adding the quantum interference of cross terms (an electron tunneling through both tips at once) from

$$\psi_1(x - u, y, 1) + \psi_1(x + u, y, 1) = \left( [\text{Cos}[x - u] + \text{Cos}[y]] + [\text{Cos}[x + u] + \text{Cos}[y]] \right) / \sqrt{e}$$

$$(17)$$

would cause additional problems. Let us examine this cross-term,

$$2\psi_1(x-u,y,1)\psi_1(x+u,y,1) = 2(Cos[x-u] + Cos[y])\,(Cos[x+u] + Cos[y])\,/\,e$$

$$= (1 + Cos[2\,u] + Cos[2\,x] + Cos[u - x - y] + Cos[u + x - y]$$

$$+ Cos[2\,y] + Cos[u - x + y] + Cos[u + x + y])\,/\,e$$

$$= (1 + Cos[2\,u] + Cos[2\,x] + 4\,Cos[u]\,Cos[x]\,Cos[y] + Cos[2\,y])\,/\,e$$

(18),

As before, we could easily Fourier Transform the simplified fourth line of (18) term by term. But given that other functions of interest are less easily reduced to a form having only one argument on which the transform will act, we will instead show the more general approach of individually transforming the two terms in the product on the first line of Eq. (18) and then convolving them.

The Fourier transform of the first term in the product in (18) is,

$$\mathcal{F}\big[Cos[x - u] + Cos[y]\big] = \pi\,\big(\delta[P']\delta[-1 + Q']$$

$$+ e^{iu}\delta[-1 + P']\delta[Q'] + e^{iu}\delta[1 + P']\delta[Q'] + \delta[P']\delta[1 + Q']\big)$$

(19).

So the Fourier convolution is of the first line in the product in (18) is,

$$\mathcal{F}\big[(Cos[x - u] + Cos[y])\,(Cos[x + u] + Cos[y])\big] = \frac{\pi^2}{2\pi}\int_{-\infty}^{\infty}dP'\int_{-\infty}^{\infty}dQ'\big(\delta[P']\delta[-1 + Q']$$

$$+ e^{iu}\delta[-1 + P']\delta[Q'] + e^{iu}\delta[1 + P']\delta[Q'] + \delta[P']\delta[1 + Q']\big)\big\{\delta[P-P']\delta[-1 + Q-Q']$$

$$+ e^{-iu}\delta[-1 + P-P']\delta[Q-Q'] + e^{-iu}\delta[1 + P-P']\delta[Q-Q'] + \delta[P-P']\delta[1 + Q-Q']\big\}$$

(20).

The first term in the parentheses in the integral, $\delta[P']\delta[-1 + Q']$, collapses the integral, replacing the bracketed terms with

$$\big\{\delta[P]\delta[-1 + Q-1] + e^{-iu}\delta[-1 + P]\delta[Q-1] + e^{-iu}\delta[1 + P]\delta[Q-1] + \delta[P]\delta[1 + Q-1]\big\}$$

and likewise for the other four terms in parentheses. Thus the Fourier Transform of the full cross term is

$$\mathcal{F}\big[2\psi_1(x - u, y, 1)\psi_1(x + u, y, 1)\big] = \frac{\pi}{e}\big\{2\,\big(1 + Cos[2\,u]\,\big)\delta[P]\,\delta[Q]$$

$$+ \big(\delta[-2 + P] + \delta[2 + P]\,\big)\delta[Q]$$

$$+ 2\,Cos[u]\,\big(\delta[-1 + P]\,\delta[-1 + Q] + \delta[1 + P]\delta[-1 + Q] + \delta[-1 + P]\,\delta[1 + Q] + \delta[1 + P]\,\delta[1 + Q]\big)$$

$$+ \delta[P]\,\big(\delta[-2 + Q] + \delta[2 + Q]\big)\big\}$$

(21).

This confirms that the conjugate lattice spacing, and hence the underlying symmetry in real space, is independent of the number of tips in the full quantum calculation, too. Again, a double tip separated by 2u introduces phase terms that suppress various frequencies at specific tip separations, but these do not always match the suppression in the classical case. At u = π/4 the points at coordinates (P,Q) = (±2,0) (lattice vectors {H,K} = ±{1,-1}) are suppressed in the classical case, but not in the quantum cross-term. Thus the full quantum tunneling current will not be so problematic at that tip separation.

At u = π/2 the point (P,Q) = (~~20~~,0) is suppressed in the quantum case but not in the classical case, the first of Equation (16), so no large affect should result there.

But in both the classical and quantum cases, u = π/2 entirely suppresses the points at (P,Q) = (±1,±1) (lattice vectors {H,K} = {±1,0} and {0, ±1}) . Furthermore at u = π/2 the classical (P,Q) = (±2,0) terms (lattice vectors {H,K} = ±{1,-1}), line 2 of Equation (16), are precisely the negative of the (P,Q) = (±2,0) quantum cross terms, line 2 of Equation (21), so we are left with only (P,Q) = (0,±2) (lattice vectors {H,K} = ±{1,1}) and (P,Q) = (0,0) nonzero at that tip separation. This is seen in Figure 7(d), where we have taken u = 1.54, just short of π/2. This will wipe out all modulation in the horizontal direction, as seen in Figure 8(d).

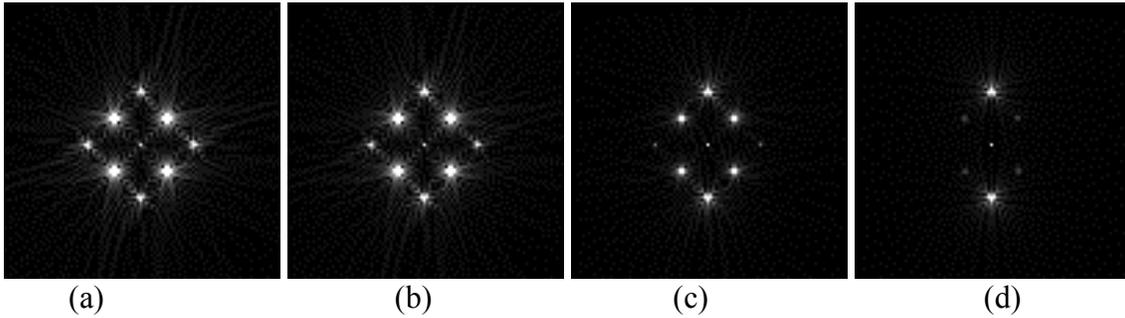

(a)         (b)         (c)         (d)

**Figure 7.** Fast Fourier transforms of the square of the quantum sum of idealized *p4mm* wave functions (15a) and (21) with (a) zero tip separation, (b) u = π/4, (c) u = 1.3, and (d) u = 1.54 = π/2 − ε.

As these Fourier spectra predict, the images of the current in this fully quantum calculation are unremarkable at u = π/4, Figure 8(b), where the classical-only calculation, Figure 5(b), showed recognizable image degradation. So, far from causing difficulties that we should avoid by arguments such as those posed by Mizes, Park, and Harrison [21], the inclusion of quantum interference *extends* the range of tip separations whose images are amenable to plane symmetry enforcement of the underlying symmetry in reciprocal space, Figure 9. The SPM tip separation limit at which we may just choose *p4mm* symmetry without a priori knowledge of the sample is about u = 1.1 for the classical-only terms Figures5(c) and 6(c) and a bit wider u = 1.3 for the fully-quantum calculation, Figures 8(c) and 9(c).

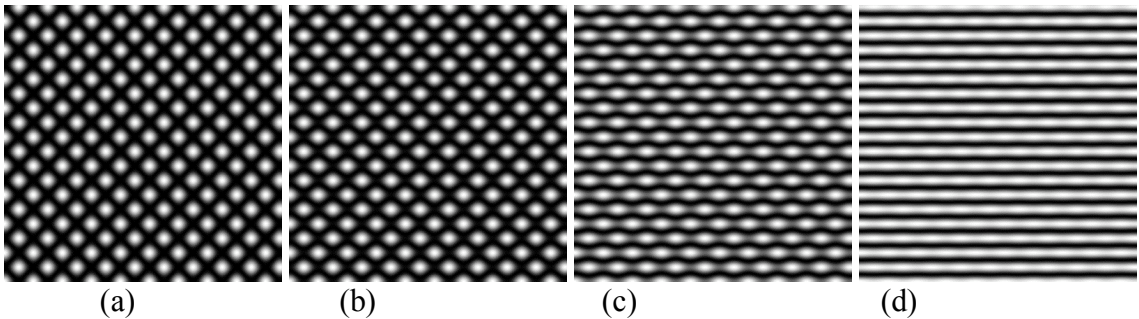

(a)         (b)         (c)         (d)

**Figure 8.** Superpositions of the two full<u>y</u>-quantum current sources with delta function STM tip half-separation (a) u = 0.0, (b) u = π/4, (c) u = 1.3, and (d) u = 1.54 = π/2 − ε.

units in the horizontal direction, respectively, relative to the *p4mm* Bloch surface functions having a period of $2\pi$.

Also as the Fourier spectrum Figure 7(d) predicts, the images of the current in this fully quantum calculation entirely lose periodicity in the horizontal direction at u = $\pi/2$, Figure 8(d). This will obviously preclude any the *p4mm* symmetry enforcement. Wherever severe effects of interference between two mini-tips change the appearance of STM images to such an extent that the CIP procedures may no longer recover the correct plane symmetry, such images may simply be discarded.

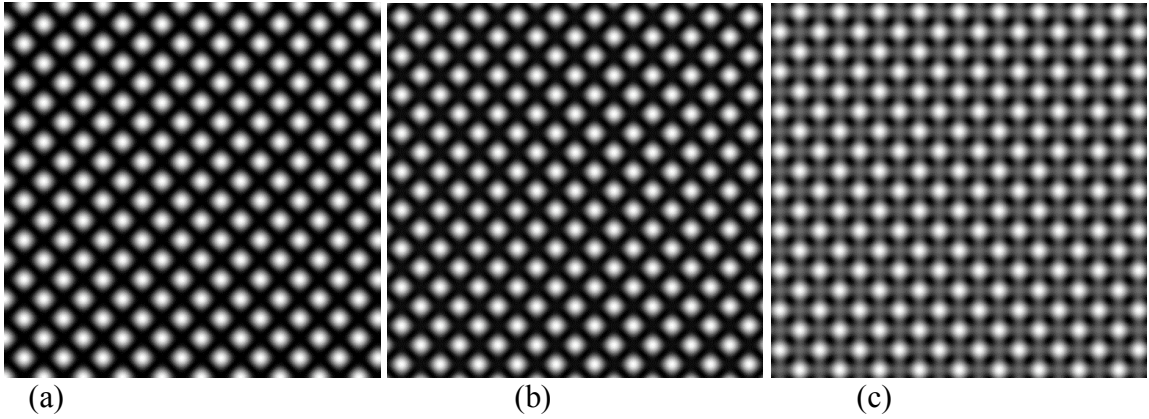

(a)                              (b)                              (c)

**Figure 9.** Plane symmetry enforcement of the underlying *p4mm* symmetry in reciprocal space for the superposition of the two classical current sources with delta function STM tip half-separation Figures 8. (a) u = 0.0, (b) u = $\pi/4$, and (c) u = 1.3 units in the horizontal direction, respectively, relative to the *p4mm* Bloch surface functions having a period of $2\pi$.

## 6. Macroscopic *Quantum Affects*

The visual difference between the classical tunneling current Figure 5(b-d) at tip half-separation u = $\pi/4 - \varepsilon$, 1.1, $\pi/2 - \varepsilon$ and the fully quantum tunneling current Figure 8(b-d) at this same tip half-separation is striking. We are tempted to suggest experimental effort to observe the periodicity of this quantum effect. Of course tip separations of 99 $\pi/4$ or even 999 $\pi/2$ would be more amenable to human manipulation than $\pi/4$ or $\pi/2$, but the present formalism predicts that quantum interference should be obtained even at such large tip separations. Indeed, that this is an example of a *macroscopic* quantum affect makes it even more interesting. Although we use a model surface wave function, it should reasonably well predict the behavior of a real *p4mm* sample.

Our highly idealized tip wave function, on the other hand, does not lend itself to confident suggestions of this sort. Thus in appendix A we refine this prediction utilizing a more realistic tip wave function, the bonding $H_2$ molecular wave function. The results, Equation (A23b), are proportional to (have the same spatial dependence as) the quantum interference results of the present section that uses the Dirac delta function double tip.

## 7. The Fourier Spectrum of Model Graphite

Consider now a somewhat more complicated atomic surface to model, highly (0001) ordered pyrolytic graphite (that is technically known as HOPG), which has three so called α sites in the hexagon that have only recently been observed experimentally by Hembacher et al. [26] using an Atomic Force Microscope (AFM) in the short range repulsive force regime, in addition to the much brighter three β sites, Figure 10.

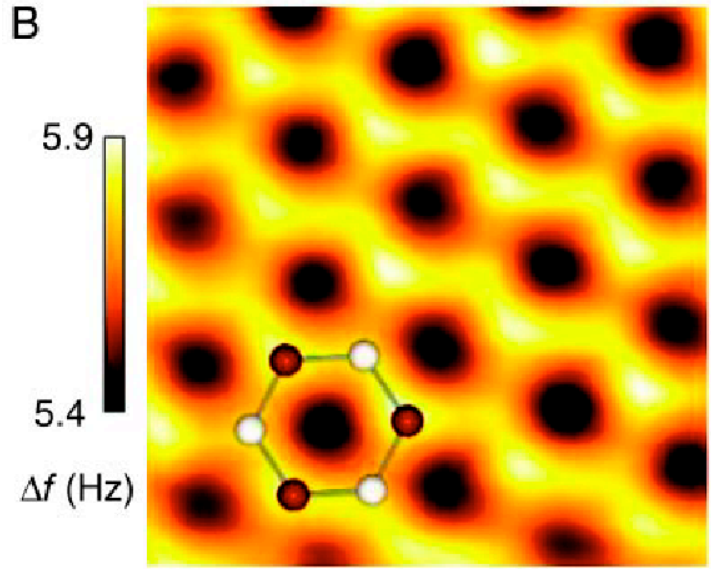

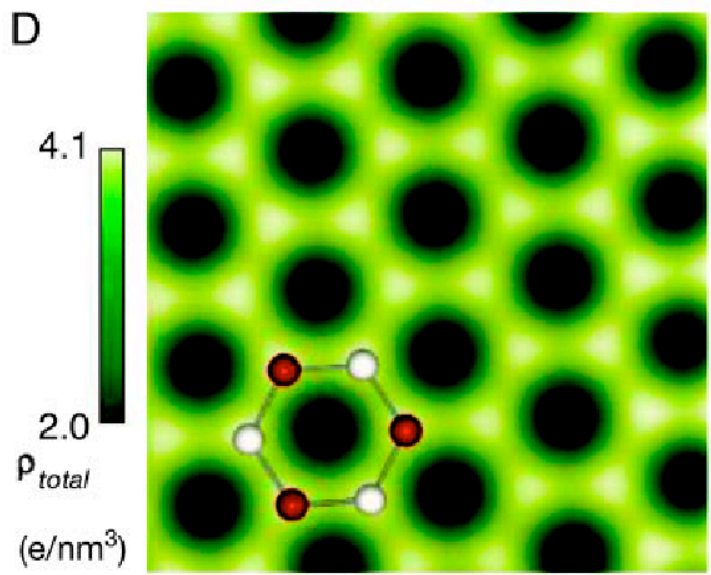

**Figure 10.** Experimental image of (0001) oriented graphite in constant-height dynamic AFM modes showing both α (as red balls) and β sites (as white ball) from Hembacher et al. [26]. The repulsive forces that are imaged in the experimental AFM image in the upper image (B) are increasing with the charge density. So the charge density calculated in the lower image (D) is a good approximation for a repulsive AFM image. The plane symmetry of these images is reasonably close to *p3m1*, i.e. the plane group that STM images of this surface should possess.

We modeled a Bloch wave function to give an idealized approximation of these graphite images. Rather than using their 12 % difference, 248 pN versus 250 pN relative to the 223 pN "hollow space" (at hexagon centers in Figure 13), we chose a wave function that provides a slightly clearer visual distinction between the two, at 28 % difference (241 versus 250 units relative to a 217 hollow space), while retaining the essential character:

$$\psi_{p3m1}^{\gamma,q}(x,y,z) = \frac{e^{-1/2\gamma z}}{5.59}\left( 100 - \cos\left(qx\right) - \cos\left[ q\left( \frac{x}{2} - \frac{\sqrt{3}y}{2} \right) \right] - \cos\left[ q\left( \frac{x}{2} + \frac{\sqrt{3}y}{2} \right) \right] \right.$$
$$\left. + \sin\left(qx\right) + \sin\left[ q\left( \frac{x}{2} - \frac{\sqrt{3}y}{2} \right) \right] + \sin\left[ q\left( \frac{x}{2} + \frac{\sqrt{3}y}{2} \right) \right] \right) \quad (22).$$

(We could, of course have made any number of other choices without changing the essential conclusions of the following work.)

As in the prior section, we first simulated "current maps" from only the superpositions of the two independent (classical) current sources. If the tips are separated in the x direction by 2u, Equation (8b) (with the real function (22)) gives such a superposed current proportional to

$$\left[ \psi_{p3m1}^{\gamma,q}(x-u,y,z) \right]^2 + \left[ \psi_{p3m1}^{\gamma,q}(x+u,y,z) \right]^2 \quad (23).$$

The Fourier transform of (23) is

$$\mathcal{F}\left[ \left[ \psi_{p3m1}^{1,1}(x-u,y,z) \right]^2 + \left[ \psi_{p3m1}^{1,1}(x+u,y,z) \right]^2 \right]$$

$$= \frac{\pi}{5.59^2 e}\left[ \left( -\text{Cos}[u/2]\left( (1584 - 1600i)\ \delta[-1 + 2P] + (1584 + 1600i)\delta[1 + 2P] \right) \right. \right.$$

$$+ (-16i)\ \text{Cos}[3u/2]\ \left( \delta[-3 + 2P] - \delta[3 + 2P] \right)\ \Big)\Big( \delta[\sqrt{3} - 2Q] + \delta[\sqrt{3} + 2Q] \Big)$$

$$+ \Big( (-2i)\ \text{Cos}[u]\ \left( \delta[-1 + P] - \delta[1 + P] \right)\ +\ 4\ \delta[P] \Big)\Big( \delta[\sqrt{3} - Q] + \delta[\sqrt{3} + Q] \Big)$$

$$+ \Big( (-2i)\ \text{Cos}[2u]\ \left( \delta[-2 + P] - \delta[2 + P] \right) + 40012\ \delta[P]$$

$$- \text{Cos}[u]\ \left( (400 - 396i)\ \delta[-1 + P] + (400 + 396i)\ \delta[1 + P] \right)\Big)\ \delta[Q]$$

$$(24).$$

As in the *p4mm* case of the previous section, the nonzero Fourier frequencies are independent of the number of tips and, hence, so is the underlying symmetry in real space. A double tip separated by 2u in the horizontal direction introduces phase terms whose only role is to suppress various frequencies at specific tip separations. When u = π/2, for instance, the points at (P,Q) = (±1,0) and (P,Q) = (±1,±$\sqrt{3}$ ) are suppressed. We may choose our lattice vector {H,K} = {1,0} to go from the origin to the upper-right point in the hexagon of Figure 11(a), whose orthogonal coordinates are $(P,Q)=\{\frac{1}{2},\frac{\sqrt{3}}{2}\}$, and {H,K} = {0,1} to go from the origin to the upper-left point in the hexagon, whose orthogonal coordinates are $(P,Q)=\{-\frac{1}{2},\frac{\sqrt{3}}{2}\}$. Then the suppressed points at rectangular coordinates (P,Q) = (±1, 0) corresponds to lattice vectors {H,K} = ±{1,-1}, seen in Figure 11(b), where we have taken u = 1.54 just short of π/2. Likewise, coordinates (P,Q) = (±1,±$\sqrt{3}$ ) correspond to lattice vectors {H,K} = {±2,0} and {0,±2}, but this next-larger ring of frequencies is not visible in the Fast Fourier transform images CRISP plots. However, numerical values appear in the corresponding frequency lists when one reduces the amplitude threshold to 80 % or less of CRISP's default value of 50. One may show that this next-larger ring terms are of order 1 % the magnitude of the inner ring, even when unsuppressed. (This is also often observed for STM images of HOPG [27]).

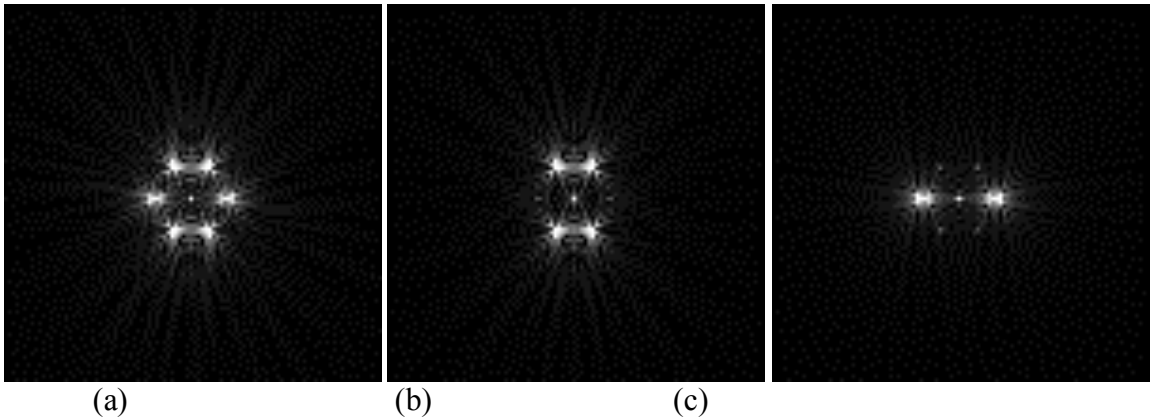

(a)                    (b)                    (c)

**Figure 11.** Fast Fourier transforms of the square of the idealized graphite wave function (22) with tip half-separation (a) u = 0.0, (b) u = 1.54=π/2−ε, and (c) u = 3.05 = π−ε units in the horizontal direction, respectively, relative to the *p3m1* Bloch surface functions having a period of 2π.

This frequency suppression results in a significant change in the "current image" registered by this model double STM tip, as seen in Figure 12(b), where we have taken the tip half-separation u = 1.54, just short of π/2.

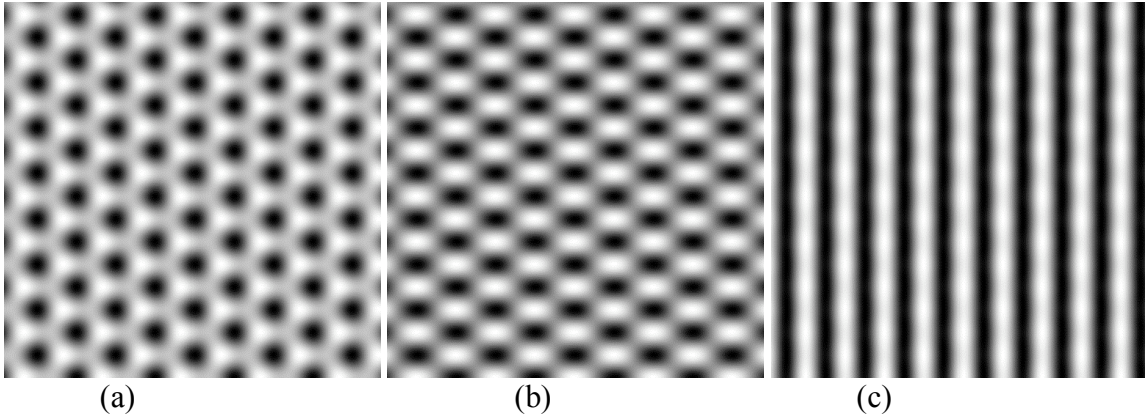

(a)              (b)              (c)

**Figure 12.** Superpositions of the two non-interfering current sources with delta function STM tip half-separation (a) u = 0.0, (b) u = 1.54 = π/2−ε, and (c) u = 3.05 = π−ε units in the horizontal direction, respectively, relative to the *p3m1* Bloch surface functions having a period of 2π.

But even with such significant suppression of frequency information, CIP still is able to recover an excellent symmetrized image of graphite, as seen in Figure 13(b), when compared with the single-tip image (a).

This is likely because of the redundancy of this three-fold symmetry: Even with all frequencies eliminated in the horizontal direction, two non-orthogonal lattice axes can still be formed through the frequencies at 60 and 120 degrees to the horizontal.

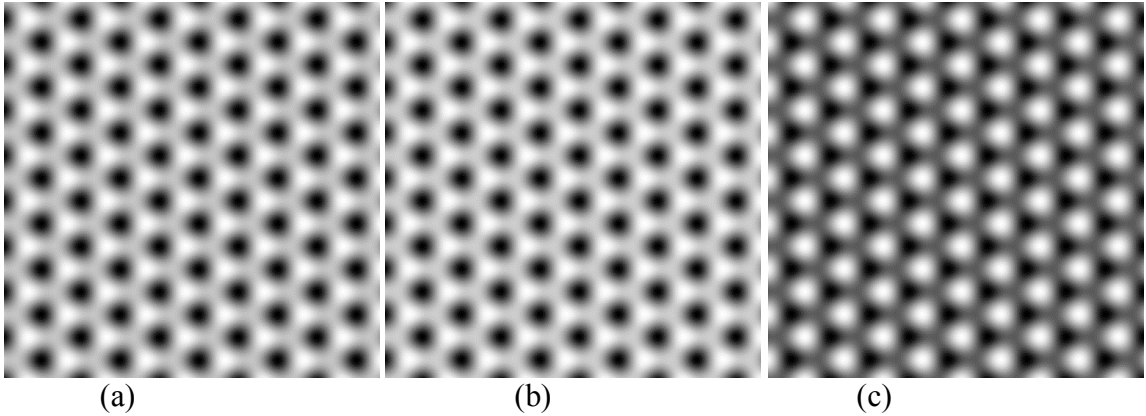

(a)              (b)              (c)

**Figure 13.** Plane symmetry enforcement of the underlying p3m1 symmetry in reciprocal space for Figure 15. (a) u = 0.0, (b) u = 1.54 = π/2−ε, and (c) u = 3.05 = π−ε units in the horizontal direction, respectively, relative to the *p3m1* Bloch surface functions having a period of 2π.

When u = π, another set of frequencies are suppressed at $(Q,P) = \{\pm\frac{1}{2}, \pm\frac{\sqrt{3}}{2}\}$ (lattice vectors {H,K} = {±1,0} and {0,±1}). This is seen in Figure 14(c), where we have taken

u = 3.05 just short of $\pi$. The suppression of higher frequencies at $(Q,P) = \{\pm\frac{3}{2}, \pm\frac{\sqrt{3}}{2}\}$, lattice vectors $\{H,K\} = \pm\{2,-1\}$ and $\mp\{1,\pm2\}$, can be observed in the HK tables CRISP generates when the amplitude threshold is reduced to 20% of its default value. Amplitudes of 61 units relative to the 9892 units of the inner ring of frequencies at u = 0 are forced to zero when u = $\pi$. The same is true for frequencies at (P,Q) = ($\pm2$,0) (or $\{H,K\} = \pm\{2,-2\}$) when u = $\pi$/4.

Such drastic suppression of frequency information along two of three possible non-orthogonal axes almost removes any periodicity in the second dimension. If one knows the "sample" has *p3m1* symmetry, CIP can recover a poor but recognizable symmetrized image of graphite, as seen in Figure 13(c), though the relative intensities of the $\alpha$ and $\beta$ sites do not match the originals (chosen to resemble Hembacher et al. [26]) well at all. Without prior knowledge of the surface, one would do well to move on to better images when faced with banding as severe as in Figure 12(c).

Next, we include the cross-term to allow for quantum interference. This will give a current proportional to

$$\left[\psi_{p3m1}^{\gamma,q}(x-u,y,z)\right]^2 + \left[\psi_{p3m1}^{\gamma,q}(x+u,y,z)\right]^2$$
$$+2\left[\psi_{p3m1}^{\gamma,q}(x-u,y,z)\right]\left[\psi_{p3m1}^{\gamma,q}(x+u,y,z)\right] = \left[\psi_{p3m1}^{\gamma,q}(x-u,y,z) + \psi_{p3m1}^{\gamma,q}(x+u,y,z)\right]^2$$

(25).

The Fourier transform of the cross terms alone is

$$\mathcal{F}\left[2\left[\psi_{p3m1}^{1,1}(x-u,y,z)\right]\left[\psi_{p3m1}^{1,1}(x+u,y,z)\right]\right]$$

$$= \frac{\pi}{5.59^2 e}\Big[+\big(16\,\text{Cos}[(3\,u)/2]\,\big(\delta[-1+2\,P] + \delta[1+2\,P]\big) + \text{Cos}[u/2]\,((-16i)\,\delta[-3+2\,P]$$

$$- (1600 - 1600i)\,\delta[-1+2\,P] - (1600+1600i)\,\delta[1+2\,P] + (16i)\,\delta[3+2\,P]\big)\big)$$

$$\times\big(\delta[\sqrt{3} - 2\,Q] + \delta[\sqrt{3} + 2\,Q]\big)$$

$$\big((-2i)\,\delta[-1+P] + 4\,\text{Cos}[u]\,\delta[P] + (2i)\,\delta[1+P]\big)\big(\delta[\sqrt{3} - Q] + \delta[\sqrt{3} + Q]\big)$$

$$+ (-2i)\,\delta[-2+P] - (4i)\,\delta[-1+P] - (400 - 400i)\,\text{Cos}[u]\,\delta[-1+P] + 40000\,\delta[P] +$$

$$+ 8\,\text{Cos}[u]\,\delta[P] + 4\,\text{Cos}[2\,u]\,\delta[P] + (4i)\,\delta[1+P] - (400+400i)\,\text{Cos}[u]\,\delta[1+P]$$

$$+ (2i)\,\delta[2+P]\big)\delta[Q]\Big]$$

(26).

The quantum cross-terms contain a small term independent of u that slightly ameliorates the complete cancellation of the frequencies at (P,Q) = ($\pm1$,0) (corresponding to lattice vectors $\{H,K\} = \pm\{1,-1\}$) at u = $\pi$/2 (and of the higher frequencies at (P,Q) = ($\pm2$,0) (or $\{H,K\} = \pm\{2,-2\}$) when u = $\pi$/4) and reduces the suppression of frequencies at (P,Q) = ($\pm1$, $\pm\sqrt{3}$ ) (or $\{H,K\} = \{\pm2, 0\}$ and $\{0,\pm2\}$) to half their u = 0 value. On the other hand, the constant-valued quantum cross-terms at frequencies at $(P,Q) = (\pm 1, \pm\sqrt{3}\,)$ precisely cancel the u-dependent classical terms at u = π.

When u = π, the quantum cross-terms also suppress frequencies at

$$(Q,P)=\{\pm\frac{1}{2},\pm\frac{\sqrt{3}}{2}\}$$ (lattice vectors {H,K} = {±1,0} and {0,±1}). (Likewise higher

frequencies are suppressed at $(P,Q)=\{\pm\frac{3}{2},\pm\frac{\sqrt{3}}{2}\}$ (or {H,K} = ±{2,-1} and ∓{1,±2})).

Overall the quantum cross terms do not alter the classical images in any discernable way.

Finally, we derive in appendix A the current "image" that a fully quantum bonding $H_2$ tip wave function [22] with significant spatial extent instead of this pair of infinitesimal Dirac delta functions would register from this sample. The results, Equation (A8b), are proportional to (have the same spatial dependence as) the quantum interference results of the present section that use the Dirac delta function double tip. This is explained by the application of Pierre Curie's well known symmetry principle. Therefore, as seen in Figure 14, this spatially extended pair of tips does not affect the outcome noted above in any observable way.

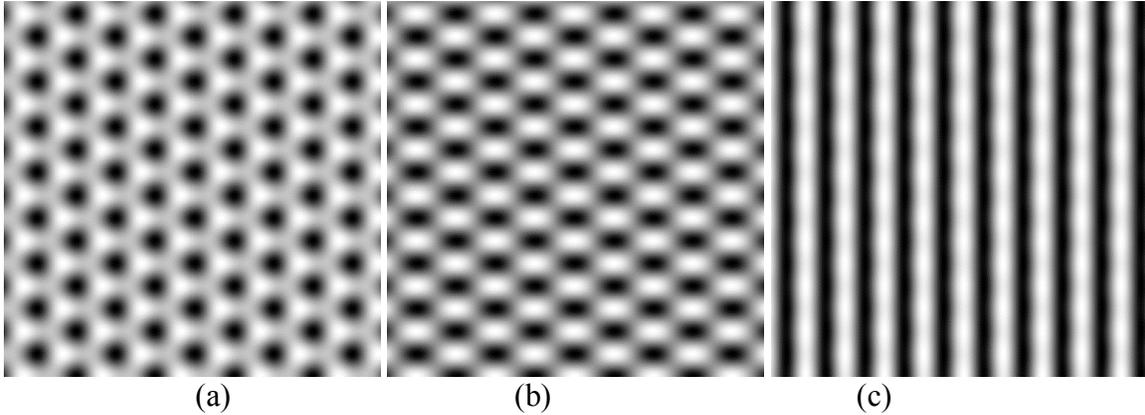

           (a)                   (b)                   (c)

**Figure 14.** Current "image" from a fully quantum bonding $H_2$ STM tip wave function with tip (inter-nuclear) half-separation (a) u = 0.0, (b) u = 1.54=π/2−ε, and (c) u = 3.05 = π−ε units in the horizontal direction, respectively, relative to the *p3m1* Bloch surface functions having a period of 2π.

## 8. Graphite with Y-axis Tip Displacement

Since the spacing between atoms in a hexagonal sample differs along the two Cartesian coordinates, for completeness we must examine interference effects in our model due to two tips displaced from each other in the vertical direction (toward the upper side of the hexagon) in addition to the horizontal direction (toward the rightmost vertex of the hexagon) discussed above.

As in the prior section, we first simulate "current maps" from only the superpositions of the two independent (classical) current sources. If the tips are separated in the y direction by 2u, Equation (8b) (with the real function (22)) gives such a superposed current proportional to

$$\left[\psi_{p3m1}^{\gamma,q}(x, y-u, z)\right]^2 + \left[\psi_{p3m1}^{\gamma,q}(x, y+u, z)\right]^2 \tag{27}.$$

The Fourier transform of (27) is

$$\mathcal{F}\left[\left[\psi_{p3m1}^{1,1}(x, y-u, z)\right]^2 + \left[\psi_{p3m1}^{1,1}(x, y+u, z)\right]^2\right]$$

$$= \frac{\pi}{5.59^2 e}\Big[\text{Cos}[\sqrt{3}\ u/2]\ \Big(\ \big(-(1584 - 1600i)\delta[-1 + 2P] - (1584 + 1600i)\delta[1 + 2P]\big)$$

$$+ (-16i)\ \big(\delta[-3 + 2P] - \delta[3 + 2P]\big)\ \Big)\big(\delta[\sqrt{3} - 2Q] + \delta[\sqrt{3} + 2Q]\big)$$

$$+ \Big((-2i)\ \text{Cos}[\sqrt{3}\ u]\ \big(\delta[-1 + P] - \delta[1 + P]\big)\ +\ 4\ \delta[P]\ \Big)\big(\delta[\sqrt{3} - Q] + \delta[\sqrt{3} + Q]\big)$$

$$+ \Big((-2i)\ \big(\delta[-2 + P] - \delta[2 + P]\big)\ -\ (400 - 396i)\ \big(\delta[-1 + P] + \delta[1 + P]\big)$$

$$+ 40012\ \delta[P]\Big)\ \delta[Q]$$

$$\tag{28}.$$

Again a double tip separated by 2u in the vertical direction introduces phase terms whose only role is to suppress various frequencies at specific tip separations. When

$u = \dfrac{\pi}{\sqrt{3}} \cong 1.81$ frequencies are suppressed at rectangular coordinates

$(Q,P) = \{\pm\dfrac{1}{2}, \pm\dfrac{\sqrt{3}}{2}\}$ (lattice vectors $\{H,K\} = \{\pm1,0\}$ and $\{0,\pm1\}$) and

$(P,Q) = \{\pm\dfrac{3}{2}, \pm\dfrac{\sqrt{3}}{2}\}$ (or $\{H,K\} = \pm\{2,-1\}$ and $\mp\{1,\pm2\}$). The former is seen in Figure

15(b), where we have taken $u = 1.7$ just short of $\pi/\sqrt{3}$ .

When $u = \dfrac{\pi}{2\sqrt{3}} \cong .9$ the frequencies are suppressed at $(P,Q) = (\pm1, \pm\sqrt{3}$ ) (or

$\{H,K\} = \{\pm2,0\}$ and $\{0,\pm2\}$) and $(P,Q) = (0,\pm\sqrt{3}$ ) (or $\{H,K\} = \pm\{1,1\}$). But the innermost ring of frequencies accounting for 99% of the amplitude override this suppression and produce a clear *p3m1* image as in Figure 16(c).

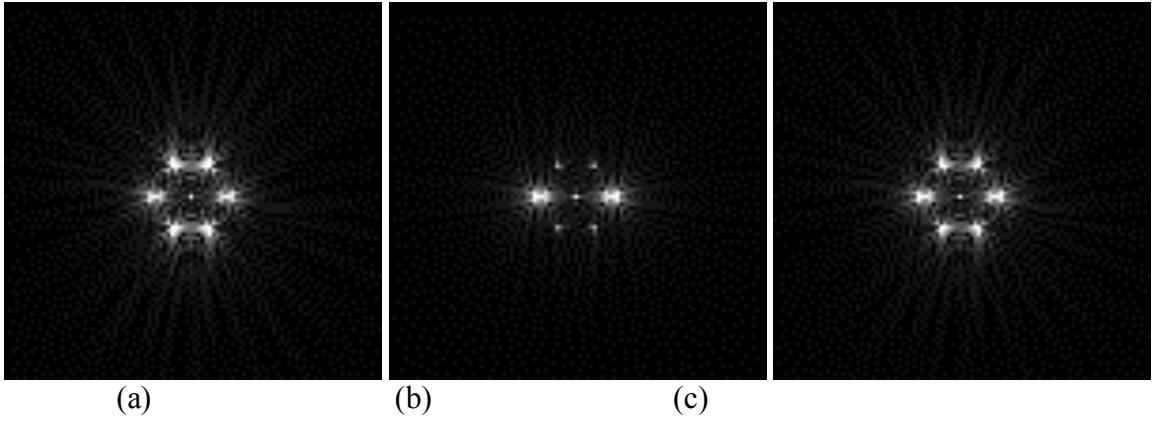

|  (a)  |  (b)  |  (c)  |

**Figure 15.** Fast Fourier transforms of the square of the idealized graphite wave function (22) with tip half-separation (a) u = 0.0, (b) u = 1.7 just short of $\pi/\sqrt{3}$, and (c) u = $\pi/\sqrt{3}/2 \approx 0.9$ units in the horizontal direction, respectively, relative to the *p3m1* Bloch surface functions having a period of $2\pi$.

The frequency suppression at u = $\frac{\pi}{\sqrt{3}} \cong 1.81$ results in a significant change in the "current image" registered by this model double STM tip, as seen in Figure 16(b), where we have taken the tip half-separation u = 1.7 just short of $\pi/\sqrt{3}$.

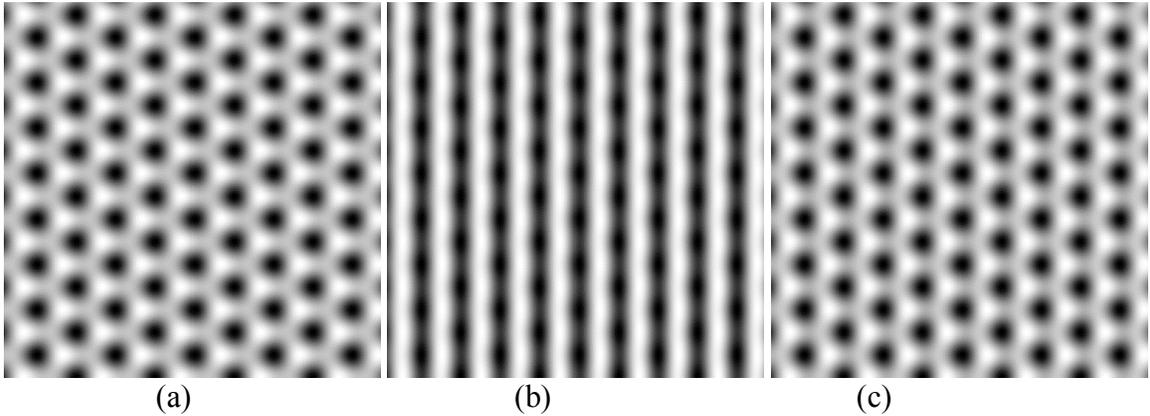

|  (a)  |  (b)  |  (c)  |

**Figure 16.** Superpositions of the two non-interfering current sources with delta function STM tip half-separation (a) u = 0.0, (b) u = 1.7 just short of $\pi/\sqrt{3}$, and (c) u = $\pi/\sqrt{3}/2 \approx 0.9$ units in the horizontal direction, respectively, relative to the *p3m1* Bloch surface functions having a period of $2\pi$.

But even with such significant suppression of frequency information, CIP still is able to recover an excellent symmetrized image of graphite, as seen in Figure 17(b), when compared with the single-tip image (a).

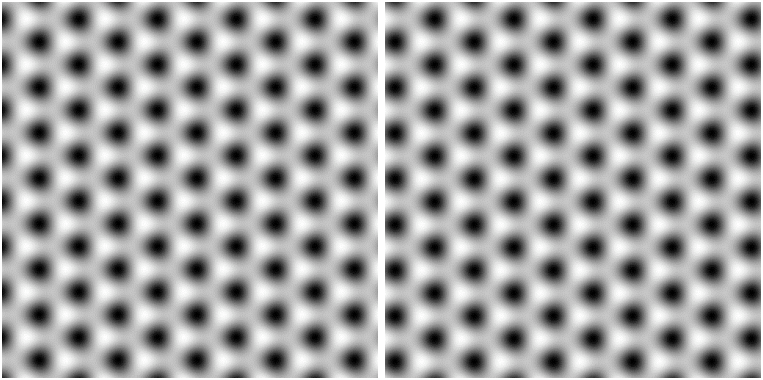

**Figure 17.** Plane symmetry enforcement of the underlying *p3m1* symmetry in reciprocal space for Figures 16. (a) u = 0.0, and (b) u = 1.7 just short of $\pi/\sqrt{3}$ units in the horizontal direction, respectively, relative to the *p3m1* Bloch surface functions having a period of 2π.

Finally, we include the cross-term that to allow for quantum interference. This will give a current proportional to

$$\left[\psi^{\gamma,q}_{p3m1}(x,y-u,z)\right]^2 + \left[\psi^{\gamma,q}_{p3m1}(x,y+u,z)\right]^2$$
$$+2\left[\psi^{\gamma,q}_{p3m1}(x,y-u,z)\right]\left[\psi^{\gamma,q}_{p3m1}(x,y+u,z)\right] = \left[\psi^{\gamma,q}_{p3m1}(x,y-u,z)+\psi^{\gamma,q}_{p3m1}(x,y+u,z)\right]^2 \quad (29).$$

The Fourier transform of the cross terms is

$$\mathcal{F}\left[2\left[\psi^{1,1}_{p3m1}(x,y-u,z)\right]\left[\psi^{1,1}_{p3m1}(x,y+u,z)\right]\right]$$

$$= \frac{\pi}{5.59^2 e}\Big[\mathrm{Cos}[\sqrt{3}u/2]\,\big((-16i)\big(\delta[-3+2P]\text{-}\delta[3+2P]\big)\text{ - }(1584-1600i)\delta[-1+2P]$$

$$\text{ - }(1584+1600i)\delta[1+2P]\,\big)\big(\delta[\mathrm{Sqrt}[3]-2Q]+\delta[\mathrm{Sqrt}[3]+2Q]\big)$$

$$+\,\big((-2i)\delta[-1+P]+4\delta[P]+(2i)\delta[1+P]\big)\big(\delta[\mathrm{Sqrt}[3]-Q]+\delta[\mathrm{Sqrt}[3]+Q]\big)$$

$$+\,\big\{\mathrm{Cos}[\sqrt{3}\,u]\big[((-4i)\big(\delta[-1+P]\text{ - }\delta[1+P]\big))+8\delta[P]\big]+(-2i)\big(\delta[-2+P]\text{ - }\delta[2+P]\big)$$

$$\text{ - }(400-400i)\delta[-1+P]+40004\delta[P]\text{ - }(400+400i)\delta[1+P]\,\big\}\delta[Q]\Big]$$

$$(30).$$

As with the classical terms, when $u = \dfrac{\pi}{\sqrt{3}} \cong 1.81$ frequencies are suppressed at rectangular coordinates $(Q,P) = \{\pm\dfrac{1}{2},\pm\dfrac{\sqrt{3}}{2}\}$ (lattice vectors {H,K} = {±1,0} and

{0,±1}) and $(P,Q) = \{\pm\dfrac{3}{2},\pm\dfrac{\sqrt{3}}{2}\}$ (or {H,K} = ±{2,-1} and ∓{1,±2}).

When $u = \dfrac{\pi}{2\sqrt{3}} \cong .9$ the frequencies are suppressed at (P,Q) = ($\pm 1$, $\pm\sqrt{3}$) (or {H,K} = {$\pm 2$,0} and {0,$\pm 2$}) and (P,Q) = (0, $\pm\sqrt{3}$) (or {H,K} = $\pm$\{1,1\}). But, as in the classical case, these are overpowered by the unsuppressed innermost ring of frequencies accounting for 99 % of the amplitude. Likewise the quantum cross terms contain unsuppressed frequencies at (P,Q) = ($\pm 1$, 0) (or {H,K} = $\pm$\{1,-1\}), that by far overshadow terms (in square brackets in the second-to-last line) at these same frequencies that are suppressed by $\pi/\sqrt{3}/2$.

### *9.* SUMMARY AND CONCLUSIONS

CIP can serve as a reliable tool for removing multiple-tip effects from SPM images in many cases. The Fourier transform step simply superposes the multiple images onto one amplitude map in Fourier space, with each contribution multiplied by a phase. We have shown that the severe banding that SPM images sometimes contain can arise from the tips approaching separations (relative to the sample's periodicity) for which those trigonometric phase function go to zero. This destroys all information along one axis of the sample surface and in such cases CIP will be unable to recover reliable surface information.

Interestingly, we find that a fully quantum model for multiple tips including interference produces a wider range of usable tip separations than a model that includes only classical image overlap. This is found to be true for tip models with and without spatial extent.

In addition, we find that quantum interference in double tips does not fall off with distance when the surface is periodic. In other words, banding that occurs at, say, $\pi/2$ will also occur at $n\,\pi/2$ for $n$ odd. We note that $n$ sufficiently large to allow for experimental manipulation of tip separation is a possible regime for observing macroscopic quantum interference.

Since our simulations were done without regards to any particulars of a STM microscope, the results of the classical simulation may be generalized to other types of SPMs as long as there are no interactions between the two tips. Also the 2D periodic surface features that a SPM images may be much larger than atomic corrugations. The results of the classical simulations can, thus, be generalized to AFM imaging.

**Acknowledgments:** This research was supported by awards from Portland State University's Venture Development Fund and the Faculty Enhancement program. A grant from Portland State University's Internationalization Council is also acknowledged.

## *Appendix A*

While the Dirac delta function double tip provides a very clear articulation of the sources of interference in image formation, it might be that a pair of tips having finite extent would result in images having entirely different characteristics. We therefore derive the current for a more realistic molecular $H_2$ tip wave function. Kobayashi and Tsukada [28] found banding in (0001) oriented graphite akin to Figure 12c using an anti-bonding $H_2$ tip wave function. We will compliment their work by using the bonding $H_2$ tip wave function that better correspond to the symmetry of the Dirac delta function double tip we have used throughout the main body of the paper.

Since the bonding $H_2$ tip wave function is already bimodal, we need only use it in the current expression for the singular tip. The quantity we seek to integrate is the quantum cross term Equation (6a).

$$A\left(\mathbf{R}, E, E'\right) = \pi^2 \int\limits_{\Omega_1} d^3 x_1 \int\limits_{\Omega_2} d^3 x_2 V_T\left(\mathbf{x}_2\right) V_T\left(\mathbf{x}_1\right) g^S\left(\mathbf{x}_1 + \mathbf{R}, \mathbf{x}_2 + \mathbf{R}, E_1\right) g^T\left(\mathbf{x}_2, \mathbf{x}_1, E_2\right)$$

(A1).

We again use our model Bloch wave function of graphite:

$$\psi_{p3m1}^{\gamma, q}(x, y, z) = \frac{e^{-1/2\gamma z}}{5.59}\left(100 - \cos\left(qx\right) - \cos\left[q\left(\frac{x}{2} - \frac{\sqrt{3}y}{2}\right)\right] - \cos\left[q\left(\frac{x}{2} + \frac{\sqrt{3}y}{2}\right)\right]\right.$$
$$\left. + \sin\left(qx\right) + \sin\left[q\left(\frac{x}{2} - \frac{\sqrt{3}y}{2}\right)\right] + \sin\left[q\left(\frac{x}{2} + \frac{\sqrt{3}y}{2}\right)\right]\right)$$

(A2)

so that the first Green's function is the one term

$$g^S\left(\mathbf{x}_1 + \mathbf{R}, \mathbf{x}_2 + \mathbf{R}, E_1\right) \equiv \sum_\sigma \psi_\sigma\left(\mathbf{x}_1 + \mathbf{R}\right)\psi_\sigma^*\left(\mathbf{x}_2 + \mathbf{R}\right)\delta\left(E_1 - E_\sigma\right)$$

$$\rightarrow \frac{e^{-1/2\gamma\left(x_{13} + R_3\right)}}{5.59}\left(100 - \cos\left[q\left(x_{11} + R_1\right)\right] - \cos\left[q\left(\frac{x_{11} + R_1}{2} - \frac{\sqrt{3}\left(x_{12} + R_2\right)}{2}\right)\right]\right.$$

$$-\cos\left[q\left(\frac{x_{11} + R_1}{2} + \frac{\sqrt{3}\left(x_{12} + R_2\right)}{2}\right)\right] + \sin\left[q\left(x_{11} + R_1\right)\right]$$

$$+ \sin\left[q\left(\frac{x_{11} + R_1}{2} - \frac{\sqrt{3}\left(x_{12} + R_2\right)}{2}\right)\right] + \sin\left[q\left(\frac{x_{11} + R_1}{2} + \frac{\sqrt{3}\left(x_{12} + R_2\right)}{2}\right)\right]\right)$$

$$\times \frac{e^{-1/2\gamma\left(x_{23} + R_3\right)}}{5.59}\left(100 - \cos\left[q\left(x_{21} + R_1\right)\right] - \cos\left[q\left(\frac{x_{21} + R_1}{2} - \frac{\sqrt{3}\left(x_{22} + R_2\right)}{2}\right)\right]\right.$$

$$-\cos\left[q\left(\frac{x_{21} + R_1}{2} + \frac{\sqrt{3}\left(x_{22} + R_2\right)}{2}\right)\right] + \sin\left[q\left(x_{21} + R_1\right)\right]$$

$$+ \sin\left[q\left(\frac{x_{21} + R_1}{2} - \frac{\sqrt{3}\left(x_{22} + R_2\right)}{2}\right)\right] + \sin\left[q\left(\frac{x_{21} + R_1}{2} + \frac{\sqrt{3}\left(x_{22} + R_2\right)}{2}\right)\right]\right)\delta\left(E_1 - E_{\gamma q}\right)$$

(A3a).

We could if we wished take sums over various values of γ and q, but we wish here to demonstrate the general character of the current image maps, not calculate precise amplitudes.

The bonding H$_2$ tip wave function [22] is

$$\psi_\tau\left(\mathbf{x}_1, \mathbf{u}\right) \rightarrow \psi_b\left(x_{11}, x_{12}, x_{13}, u\right) = e^{-\sqrt{\left(x_{11} - u\right)^2 + x_{12}^2 + x_{13}^2}} + e^{-\sqrt{\left(x_{11} + u\right)^2 + x_{12}^2 + x_{13}^2}} \qquad \text{(A4a)}.$$

Then

$$g^T\left(\mathbf{x}_2, \mathbf{x}_1, E_2\right) \equiv \sum_\tau \psi_\tau\left(\mathbf{x}_2\right)\psi_\tau^*\left(\mathbf{x}_1\right)\delta\left(E_2 - E_\tau\right) \rightarrow \left(e^{-r_{2-}} + e^{-r_{2+}}\right)\left(e^{-r_{1-}} + e^{-r_{1+}}\right)\delta\left(E_2 - E_{ab}\right)$$

(A5a),

where

$$r_{j\mp} = \sqrt{\left(x_{j1} \mp u\right)^2 + x_{j2}^2 + x_{j3}^2} \qquad \text{(A6)}.$$

With these wave functions the current (A1) becomes

$$A\left(\mathbf{R}, E, E'\right) = D_b^2\left(\mathbf{R}, \mathbf{u}\right) \qquad \text{(A7)},$$

where

$$D_b(\mathbf{R},u) = \pi V_T \int_{\Omega_T} d^3x \, \frac{e^{-1/2\gamma(x_3+R_3)}}{5.59} \left( 100 - \cos\left[ q(x_1+R_1) \right] - \cos\left[ q\left( \frac{x_1+R_1}{2} - \frac{\sqrt{3}(x_2+R_2)}{2} \right) \right] \right.$$

$$-\cos\left[ q\left( \frac{x_1+R_1}{2} + \frac{\sqrt{3}(x_2+R_2)}{2} \right) \right] + \sin\left[ q(x_1+R_1) \right]$$

$$\left. + \sin\left[ q\left( \frac{x_1+R_1}{2} - \frac{\sqrt{3}(x_2+R_2)}{2} \right) \right] + \sin\left[ q\left( \frac{x_1+R_1}{2} + \frac{\sqrt{3}(x_2+R_2)}{2} \right) \right] \right] \left[ e^{-r_-} + e^{-r_+} \right]$$

$$= \left[ S_{c-}(\mathbf{R},u) + S_{c-}(\mathbf{R},u) \right] - \left[ C_{1-}(\mathbf{R},u) + C_{1-}(\mathbf{R},u) \right] - \left[ C_{1\text{-}2-}(\mathbf{R},u) + C_{1\text{-}2-}(\mathbf{R},u) \right]$$

$$- \left[ C_{1+2-}(\mathbf{R},u) + C_{1+2-}(\mathbf{R},u) \right] + \left[ S_{1-}(\mathbf{R},u) + S_{1-}(\mathbf{R},u) \right]$$

$$+ \left[ S_{1\text{-}2-}(\mathbf{R},u) + S_{1\text{-}2-}(\mathbf{R},u) \right] + \left[ S_{1+2-}(\mathbf{R},u) + S_{1+2-}(\mathbf{R},u) \right]$$

<div align="right">(A8a),</div>

in which we have dropped mention of the energy conservation coefficients for notational convenience, and take the integration region to be inside the region of constant potential. The brackets cluster one surface term multiplying each of the two tip terms in order, with the final indices referring to the terms with $e^{-r_-}$ and $e^{-r_+}$ respectively. The numbered indices refer to the trigonometric function including $x_1$, or $x_2$ or both, and the middle sign is that of the $x_2$ term of the trigonometric function.

Evaluating integrals that have a coordinate dependence with the specific complication of the bonding $H_2$ wave function are often carried out using hyper-spherical coordinates [29], Jacobi coordinates or bond-length coordinates [30], or confocal ellipsoidal coordinates [31]. We find it far easier to slightly modify a strategy developed by one of us (Straton) [32] for evaluating integrals of any number of products of multicenter ground-state or excited [33] atomic wave functions, Coulomb or Yukawa potentials, and Coulomb-waves [34].

The problem those techniques solved is to segregate out the various coordinate components of multicenter wave functions and potentials. The simplest case is the overlap integral

$$(-1)^m S_{n\ell-m,n'\ell'm'}^{\lambda,\lambda'}(0;\mathbf{R}) = \int d\mathbf{r} \, \psi_{n\ell m}^{\lambda*}(\mathbf{r}) \psi_{n'\ell'm'}^{\lambda'}(\mathbf{r}-\mathbf{R}) \qquad (A9).$$

In the spherically-symmetric case [35] of a hydrogenic ground-state, for instance, each wave function is Gaussian transformed [36]

$$\psi_{1s}^{\lambda}(\mathbf{x}) = \frac{\lambda^{5/2}}{2\pi} \int_0^{\infty} d\rho \, \frac{e^{-x^2\rho-\lambda^2/4\rho}}{\rho^{3/2}}, \qquad x \geq 0, \; \lambda > 0 \qquad (A10a).$$

Then

$$S_{1s,1s}^{\lambda,\lambda'}(0;\mathbf{R}) = \frac{(\lambda\lambda')^{5/2}}{(2\pi)^2} \int_0^{\infty} d\rho_1 \, \frac{e^{-\lambda^2/4\rho_1}}{\rho_1^{3/2}} \int_0^{\infty} d\rho_2 \, \frac{e^{-\lambda^2/4\rho_2}}{\rho_2^{3/2}} \int d\mathbf{r} \, e^{-r^2\rho_1-(\mathbf{r}-\mathbf{R})^2\rho_2} \qquad (A11).$$

One can complete the square in the quadratic form in the exponential and do the $\mathbf{r}$ integral, followed by the $\rho$ integrals. For very complex integrals, completing the square in the quadratic forms appears to be an arduous task. Fortunately, the orthogonal transformation that accomplishes this task need not be explicitly found since only its invariant determinant is left after the spatial integration is done. Thus the final form of the integral may be found in the general case, [32, 33, 34] and one may just plug the specifics into the final result.

In the present case we cannot simply add the specifics and use the final form because the $x_3$ integrals are done on the interval $[-b, \infty]$ not $[-\infty, \infty]$ and the coefficients in the quadratic form have a slightly different structure. But the present integral is sufficiently simple that we can just Gaussian transform the bonding $H_2$ wave function term to facilitate direct integration over x,

$$e^{-\lambda_3 r_{\mp}} = \frac{\lambda_3}{2\sqrt{\pi}} \int_0^\infty d\rho_3 \frac{e^{-r_{\mp}^2 \rho_3 - \lambda_3^2/4\rho_3}}{\rho_3^{3/2}}, \qquad r_{\mp} \geq 0, \ \lambda > 0$$

(A10b).

Consider first the third and fourth terms of (A8a),

$$C_{1\mp}(\mathbf{R}, u) = \frac{\pi V_T}{5.59} \frac{\lambda_3}{2\sqrt{\pi}} \int_{\Omega_T} d^3x \, e^{-1/2\gamma(x_3 + R_1)} \cos\left[q(x_1 + R_1)\right] \int_0^\infty d\rho_3 \frac{e^{-\left[(x_1 \mp u)^2 + x_2^2 + x_3^2\right]\rho_3 - \lambda_3^2/4\rho_3}}{\rho_3^{3/2}}$$

(A12).

The integration volume can be taken as any volume that encloses the tip, the region to the right of $S_0$ of Equation (2b) and Figure 2. It is convenient to take $S_0$ as an infinite plane parallel to the sample surface. Then the $x_2$ integral, parallel to the surface, has value [37]

$$2\int_0^\infty dx_2 e^{-\rho_3 x_2^2} = \frac{\sqrt{\pi}}{\rho_3^{1/2}}, \qquad\qquad \rho_3 > 0$$

(A13).

For the $x_1$ integral, also parallel to the surface, we change variables to

$$x_1' = x_1 \mp u$$

(A14),

with unit Jacobian, and [38]

$$\int_{-\infty}^\infty dx_1' e^{-\rho_3 x_1'^2} \cos\left[q(x_1' \pm u + R_1)\right] = \frac{\sqrt{\pi}}{\rho_3^{1/2}} \, e^{-q^2/4\rho_3} \cos\left[q(R_1 \pm u)\right]$$

(A15a).

We have taken $x_3 = 0$ on the axis of the bonding $H_2$ wave function so that $x_3$ in the integral takes on negative values down as close to the sample surface as we choose, a distance we will call $b$. Rather than completing the square in $x_3$ and having to deal with an integration variable in the $\rho_3$ lower limit, it is cleaner to first do the $\rho_3$ integral [39]:

$$\int_0^\infty d\rho_3 \, \rho_3^{-2-1/2} e^{-x_3^2 \rho_3 - \mu/4\rho_3} = \frac{\sqrt{\pi} e^{-\mu^{1/2} x_3}}{2(\mu/4)^{3/2}} \left(1 + \mu^{1/2} x_3\right), \qquad \text{Re } \mu > 0, \ \text{Re } x_3^2 > 0$$

(A18),

where

$$\mu = \lambda_3^2 + q^2$$

(A19a)

in which $\lambda_3 = 1$. Finally [40],

$$C_{1\mp}(\mathbf{R},u) = \frac{2\pi^2 V_T}{5.59} \frac{\lambda_3}{\mu^{3/2}} e^{-1/2\gamma R_3} \cos\left[q\left(R_1 \pm u\right)\right] \int\limits_{-b}^{\infty} dx_3 \left(1 + \mu^{1/2} x_3\right) e^{-\left(\mu^{1/2} + \gamma/2\right) x_3}$$

$$= \frac{2\pi^2 V_T}{5.59} \frac{\lambda_3}{\mu^{3/2}} e^{-1/2\gamma R_3} \cos\left[q\left(R_1 \pm u\right)\right] \times$$

$$\times \left[ e^{-\left(\mu^{1/2} + \gamma/2\right) x_3} \left( \frac{1}{-\left(\mu^{1/2} + \gamma/2\right)} + \mu^{1/2} \left\{ \frac{x_3}{-\left(\mu^{1/2} + \gamma/2\right)} - \frac{1}{\left(\mu^{1/2} + \gamma/2\right)^2} \right\} \right) \right]_{-b}^{\infty}$$

$$= \frac{2\pi^2 V_T}{5.59} \frac{\lambda_3}{\mu^{3/2}} e^{\left(\mu^{1/2} + \gamma/2\right) b} \left( \frac{2\mu^{1/2} + \gamma/2}{\left(\mu^{1/2} + \gamma/2\right)^2} - \frac{\mu^{1/2} b}{\left(\mu^{1/2} + \gamma/2\right)} \right) e^{-1/2\gamma R_3} \cos\left[q\left(R_1 \pm u\right)\right]$$

$$\equiv C_{1\mp}(\mathbf{R},u,\gamma,q,\lambda_3)$$

(A20a),

where we put all of the dependent variables in the final line, but will generally use the shortened version.

One may likewise show that

$$S_{1\mp}(\mathbf{R},u) = C_{1\mp}(\mathbf{R},u) \frac{\sin\left[q\left(R_1 \pm u\right)\right]}{\cos\left[q\left(R_1 \pm u\right)\right]}$$

(A20b),

$$S_{1\sigma 2\mp}(\mathbf{R},u) \equiv S_{1\mp}(\mathbf{R},u,\gamma,q,\lambda_3) = C_{1\mp}(\mathbf{R},u,\gamma,q,\lambda_3) \frac{\sin\left[q\left(\frac{R_1 \pm u}{2} + \sigma\frac{\sqrt{3}R_2}{2}\right)\right]}{\cos\left[q\left(R_1 \pm u\right)\right]}$$

(A20c),

and

$$C_{1\sigma 2\mp}(\mathbf{R},u) = C_{1\mp}(\mathbf{R},u) \frac{\cos\left[q\left(\frac{R_1 \pm u}{2} + \sigma\frac{\sqrt{3}R_2}{2}\right)\right]}{\cos\left[q\left(R_1 \pm u\right)\right]}$$

(A20d),

where in the latter two cases we apply Equation (A15a) for both $x_1$ and $x_2$ and multiply the resulting exponentials to get $e^{-q^2/4^2\rho_3} e^{-3q^2\sigma^2/4^2\rho_3} = e^{-q^2/4\rho_3}$ again.

For the first two constant terms, we apply Equation (A13) for both $x_1$ and $x_2$ and thus there is no q-dependence in (A19a):

$$S_{c\mp}(\mathbf{R},u,\gamma,q,\lambda_3) = S_{1\mp}(\mathbf{R},u,\gamma,0,\lambda_3)$$

(A21),

This means that they simply add to contribute a constant term in (A8a).

Finally, the current image amplitude is

$$D_b(\mathbf{R}, u) = \frac{2\pi^2 V_T \lambda_3 e^{-1/2 \gamma R_3}}{5.59} \left\{ \frac{e^{(\mu^{1/2} + \gamma/2)b}}{\mu^{3/2}} \left( \frac{2\mu^{1/2} + \gamma/2}{\left(\mu^{1/2} + \gamma/2\right)^2} - \frac{\mu^{1/2}b}{\left(\mu^{1/2} + \gamma/2\right)} \right) \right.$$

$$\left( [100 + 100] - \left[ \cos\left[q\left(R_1 - u\right)\right] + \cos\left[q\left(R_1 + u\right)\right] \right] \right)$$

$$- \left[ \cos\left[ q\left( \frac{R_1 - u}{2} - \frac{\sqrt{3}R_2}{2} \right) \right] + \cos\left[ q\left( \frac{R_1 + u}{2} - \frac{\sqrt{3}R_2}{2} \right) \right] \right]$$

$$- \left[ \cos\left[ q\left( \frac{R_1 - u}{2} + \frac{\sqrt{3}R_2}{2} \right) \right] + \cos\left[ q\left( \frac{R_1 + u}{2} + \frac{\sqrt{3}R_2}{2} \right) \right] \right]$$

$$+ \left[ \sin\left[ q\left( R_1 - u \right) \right] + \sin\left[ q\left( R_1 + u \right) \right] \right]$$

$$+ \left[ \sin\left[ q\left( \frac{R_1 - u}{2} - \frac{\sqrt{3}R_2}{2} \right) \right] + \sin\left[ q\left( \frac{R_1 + u}{2} - \frac{\sqrt{3}R_2}{2} \right) \right] \right]$$

$$\left. + \left[ \sin\left[ q\left( \frac{R_1 - u}{2} + \frac{\sqrt{3}R_2}{2} \right) \right] + \sin\left[ q\left( \frac{R_1 + u}{2} + \frac{\sqrt{3}R_2}{2} \right) \right] \right] \right\}$$

(A8b).

The current images for the square (A7) of this amplitude are shown in Figure 14 of the main part of the paper. There is no discernable difference in the current maps for this extended double-tip and the current maps for a pair of Dirac delta function tips, Figure 12. Indeed, this has the identical **R**-dependence as the quantum Dirac delta function tip result $\left( \left[ \psi_{p3m1}^{\gamma, q}(R_1 - u, R_2, R_3) \right] + \left[ \psi_{p3m1}^{\gamma, q}(R_1 + u, R_2, R_3) \right] \right)^2$ (see Equation (25)).

### *The p4mm case*

With these calculations done, we may immediately apply them to the case of Chen's [23] *p4mm* cosine Bloch wave function of Equation (14),

$$\psi_1(x, y, z) = e^{-1/2 \gamma z}\left( \cos(qx) + \cos(qy) \right)$$

(A22).

The amplitude is then

$$D_t(\mathbf{R}, u) = \pi V_T \int_{\Omega_T} d^3x\, e^{-1/2 \gamma (x_3 + R_3)} \left( \cos\left[ q\left( x_1 + R_1 \right) \right] + \cos\left[ q\left( x_2 + R_2 \right) \right] \right)\left( e^{-r_-} + e^{-r_+} \right)$$

$$= 5.59\left\{ \left[ C_{1-}(\mathbf{R}, u) + C_{1-}(\mathbf{R}, u) \right] + \left[ C_{2-}(\mathbf{R}, u) + C_{2-}(\mathbf{R}, u) \right] \right\}$$

(A23a).

Its third and fourth terms are:

$$C_{2\mp}(\mathbf{R},u) = C_{1\mp}(\mathbf{R},u) \frac{\cos[qR_2]}{\cos[q(R_1 \pm u)]} \tag{A24}.$$

Then

$$\begin{aligned}
D_t(\mathbf{R},u) = 2\pi^2 V_T \lambda_3 &\left\{ \frac{e^{(\mu^{1/2}+\gamma/2)b}}{\mu^{3/2}} \left( \frac{2\mu^{1/2}+\gamma/2}{(\mu^{1/2}+\gamma/2)^2} - \frac{\mu^{1/2}b}{(\mu^{1/2}+\gamma/2)} \right) \right. \\
&\times e^{-1/2\gamma R_3} \left[ \cos[q(R_1-u)] + \cos[q(R_1+u)] + 2\cos[qR_2] \right] \\
&\left. \sim \Big[ \big(\cos[q(R_1-u)] + \cos[qR_2]\big) + \big(\cos[q(R_1+u)] + \cos[qR_2]\big) \Big] \right\}
\end{aligned}$$
$$(A23b),$$

which is clearly proportional to the fully quantum calculation using the Dirac delta function tip. This validates our use of that tip for the sake of clarity.

Note that these results are to be expected for both kinds of wavefunctions from Pierre Curie's well know symmetry principle: *"c'est la dissymmétrie qui crée le phénoméne"* (it's dissymmetry that creates the phenomenon) [41].

Applied to STM imaging and ignoring noise, Curie's principle states that the recorded local electronic density of states at the Fermi level is the symmetry group intersection of the local site symmetries with the point symmetry of the scanning probe tip. The point symmetries of a double-Dirac delta function- tip and the bonding $H_2$ tip are the same so that formally identical STM images must be obtained for the same scanning direction of both tips.

# Appendix B

In order to determine the plane symmetry to which an image most likely belongs one may define Fourier coefficient amplitude ($A_{res}$) and phase angle ($\varphi_{res}$) residuals as

$$A_{res} = \frac{\sum\limits_{H,K} \left\| A_{obs}(H,K) \right| - \left| A_{sym}(H,K) \right\|}{\sum\limits_{H,K} \left| A_{obs}(H,K) \right|} \qquad \text{in percentage} \tag{B1}$$

and

$$\varphi_{res} = \frac{\sum\limits_{H,K} w(H,K) \cdot \left| \varphi_{obs}(H,K) - \varphi_{sym}(H,K) \right|}{\sum\limits_{H,K} w(H,K)} \qquad \text{in degrees} \tag{B2},$$

where the subscripts stand for *obs*erved and *sym*metrized, $w$ is a relative weight (that is set proportional to $A_{obs}$), and the sums are taken over all Fourier coefficient labels H and K, refs. [42,43]. For the plane symmetry of an image to be close to a certain plane group, both of these residuals need to be small. A small phase residual is generally more useful as an indication for the likelihood to which plane group an image most likely belongs.

Higher symmetric plane groups (such as *p4mm*) possess a higher multiplicity of the general position per primitive cell than their type I *("translationsgleiche")* subgroups (such as *p4*, *c2mm* and *p2mm*). These subgroups result from their type I supergroup by the removal of certain symmetry elements [11]. Whenever the Fourier coefficient residuals of (B1) and (B2) of an image are for a higher symmetric plane group not significantly larger than for its respective type I subgroups, one has to conclude that this particular group is the more likely plane group.

In addition, we utilize the $A_{forbidden}/A_{allowed}$ ratio for those six plane groups that possess systematic absences [11]. This ratio is defined as the amplitude sum of the Fourier coefficients that are forbidden by the plane symmetry but were nevertheless observed ($A_{forbidden}$) divided by the amplitude sum of all other observed Fourier coefficients that are allowed ($A_{allowed}$) by the plane symmetry. If this ratio is larger than zero, deviations from the respective plane symmetry are actually present and the plane group forbidden Fourier transform coefficients have amplitudes greater than zero. For the six plane groups to which this ratio is applicable, a large ratio makes it more unlikely that the respective group is the most likely plane symmetry group that a 2D-periodic image may possess when geometric distortions, defects and noise are removed.

Note that systematic absences are due to both glide lines and unit cell centering [11]. The $A_{forbidden}/A_{allowed}$ ratio ($A_f/A_a$ for short below), thus, helps one decide whether an image is closer to plane groups *pm* or *p2mm* than to plane groups *cm* or *c2mm*, respectively. One, thus, possesses a set of quantitative measures, i.e. the combination of this ratio with the Fourier coefficient amplitude and phase angle residuals B1 and B2), that can help one decide which plane group an image most likely possesses.

After these introductory paragraphs on the plane symmetry group determination methodology, we are now ready to assess the double tip images of Figure 5.

For u = 0.77 =$\pi/4-\varepsilon$, Figure 5b, we obtain with CRISP for the non-hexagonal plane groups the following standard plane symmetry quantifiers, Table 1.

Table 1: Standard plane symmetry quantifiers obtained from the application of CRISP to Figure 5(b).

|  | p2 | p1m1 | p11m | p1g1 | p11g | c1m1 | c11m | p2mm | p2mg | p2gg | c2mm | p4 | p4mm | p4gm |
|---|---|---|---|---|---|---|---|---|---|---|---|---|---|---|
| $A_{res}$ | n.d. | 25.7 | 25.7 | 40.4 | 40.4 | 94.7 | 94.7 | 25.7 | 40.4 | 40.4 | 94.7 | 94.7 | 25.8 | 25.8 | 94.7 |
| $\varphi_{res}$ | 0.2 | 0.3 | 0.2 | 2.1 | 2.1 | 0.0 | 0.0 | 0.2 | 1.9 | 2.0 | 0.2 | 0.2 | 0.2 | 0.2 |
| $A_f/A_a$ | n.d. | n.d. | n.d. | 1.0 | 1.0 | 0 | 0 | n.d. | 1.0 | 1.0 | 0 | n.d. | n.d. | n.d. |

From this table, we can conclude confidently that *p4mm* is the most likely plane group, since it must satisfy twice as many symmetry constraints as *p2mm* (due to the higher multiplicity of its general position) while possessing similarly low residuals. The enforcement of *p4mm* results in the removal of the classical double tip effect, Figure 6(b).

For a tip half-separation of u = 1.1, Figure 5(c), the visual "basket weave" structure of the image suggests a more severe breaking of the *p4mm* symmetry. This is indeed observed in the respective table of standard plane symmetry quantifiers for this image, Table 2.

(The *c2mm* amplitude residual, for instance, has dropped from four times as large as the *p4mm* residual in Table 1 to twice as large in Table 2.) The data nevertheless lead to the conclusion that *p4mm* is the most likely plane symmetry group for this image. The zero $A_f/A_a$ ratios are a byproduct of the very low number of Fourier coefficients in Figure 5(c) and are, therefore, not particularly useful in this special case.

Table 2: Standard plane symmetry quantifiers obtained from the application of CRISP to Figure 5(c).

| | p2 | p1m1 | p11m | p1g1 | p11g | c1m1 | c11m | p2mm | p2mg | p2gm | p2gg | c2mm | p4 | p4mm | p4gm |
|---|---|---|---|---|---|---|---|---|---|---|---|---|---|---|---|
| $A_{res}$ | n.d. | 13.7 | 13.7 | 18.6 | 18.6 | 28.7 | 28.7 | 13.7 | 18.6 | 18.6 | 28.9 | 28.7 | 13.8 | 13.9 | 29.0 |
| $\varphi_{res}$ | 0.5 | 0.3 | 0.3 | 31.6 | 31.6 | 0.1 | 0.1 | 0.7 | 32.0 | 31.8 | 0.3 | 0.3 | 0.5 | 0.5 | 0.3 |
| $A_f/A_a$ | n.d. | n.d. | n.d. | 0 | 0 | 0 | 0 | n.d. | 0 | 0 | 0 | 0 | n.d. | n.d. | 0 |

For a (double-tip) half separation of u = 1.54 = $\pi/2 - \varepsilon$, Figure 5(d), on the other hand, we have passed the limit of where we could choose plane symmetry *p4mm* from an image that is considerably obscured by classical superposition of two images. (That limit, thus, seems to be a tip half-separation of about u = 1.1, Figures 5(c) and 6(c).) This becomes evident from the data of Table 3 as obtained with CRISP.

Table 3: Standard plane symmetry quantifiers obtained from the application of CRISP to Figure 5(d).

| | p2 | p1m1 | p11m | p1g1 | p11g | c1m1 | c11m | p2mm | p2mg | p2gm | p2gg | c2mm | p4 | p4mm | p4gm |
|---|---|---|---|---|---|---|---|---|---|---|---|---|---|---|---|
| $A_{res}$ | n.d. | 0.1 | 0.1 | 0.1 | 0.1 | 0.0 | 0.0 | 0.1 | 0.1 | 0.1 | 0.1 | 0.0 | 0.1 | 0.1 | 0.2 |
| $\varphi_{res}$ | 0.5 | 2.8 | 2.5 | 0.2 | 0.8 | 0.1 | 0.2 | 5.4 | 0.8 | 0.3 | 0.8 | 0.2 | 5.4 | 5.4 | 0.8 |
| $A_f/A_a$ | n.d. | n.d. | n.d. | 1.0 | 1.0 | 0 | 0 | n.d. | 1.0 | 1.0 | 0 | 0 | n.d. | n.d. | 0 |

From this table, we have to conclude that *c2mm* is the most likely plane group of the image of Figure 5(d). It, therefore, does not make sense to symmetrize this image to *p4mm*. Indeed, the (type I, "translationsgleiche" [11]) *c2mm* subgroup of *p4mm* can be considered as arising from a "non-uniform scaling" of the site symmetries of the periodic motif along the diagonal of the quadratic unit cell of *p4mm* (whereby the unit cell translations remain invariant).

As one may glean from Figure 5, the motifs of (b) and (c) are indeed progressively "scaled" along the diagonal direction of the quadratic unit cell of *p4mm* in (a) At some point of the scaling process beyond u = 1.1, the transition to plane group *c2mm* must necessarily occur. This visual comprehension of the progressive changes in Figure 5 is also borne out by the plane symmetry quantifiers of Tables 1 to 3 above. With progressive non-uniform scaling of the periodic motif due to an increasing separation of the two tips, the quantifiers for *c2mm* and its type I subgroups *c1m1* and *c11m* are falling off on a relative scale (and also an absolute scale) while the quantifiers of *p4mm* and its type I subgroups *p4* and *p2mm* are increasing on a relative scale.

Finally, one can consider how selected site symmetries of plane group *p4mm* either change with or remain invariant to such a non-uniform scaling. The site symmetries of a periodic motif in plane group *p4mm* that are reduced by a non-uniform scaling due to the effects of a double tip scanning in the [1,1] direction are obviously *4mm* at positions (0,0) and (0.5,0.5). These site symmetries simply reduce to *2mm* by way of removal of one set of mirror lines. The site symmetries ..*m* at positions (x,x) and (-x,x), on the other hand, remain but form after this scaling mirror lines along the unit cell axes of *c2mm*. All 4 two-fold rotation points and the 2 glide lines of *p4mm* are only affected by the scaling in so far as that they acquire coordinates that are compliant with the *c2mm* symmetry.

Henderson, J. M. Smith, MRC image processing programs, J. Struct. Biol. 116 (1996) 9-16).

Crystallographic image processing routines are currently programmed in MATLAB at Portland State University (PSU) by Mr. Taylor Bilyeu. The aim of that project is to support crystallographic image processing from scanning probe microscopy images. The functionality of this program will, therefore, be somewhat reduced with respect to the quasi-standards of electron crystallography (but include options that are SPM image processing specific, e.g. a Wiener filter for noise reduction and difference Fourier syntheses. PSU's program will also include the extraction of the prevalent point spread function of a SPM from an image of a highly symmetric 2D periodic calibration sample and the usage of the inverse of this function for the "quasi-needle symmetrization" of the prevailing scanning probe tip in images of unknowns that were recorded under the same experimental conditions. A demonstration version of this program will be made available at: http://nanocrystallography.research.pdx.edu and its mirror site: http://nanocrystallography.net. The compiled code and/or source code will be available over PSU's Office for Graduate Research and Sponsored Projects.

Ryzhik, *Table of Integrals, Series, and Products* 5ed (Academic, New York, 1994), p. 385, No. 3.471.12, with the additional limit $\left| Arg \left( q/x_3 \right)^2 \right| < \pi/2$ .